\documentclass[paper,11pt]{JHEP}
\usepackage[centertags]{amsmath}
\usepackage{amsfonts} \usepackage{amssymb} \usepackage{amsthm}
\allowdisplaybreaks[1]
\usepackage{graphicx}
\newcommand{\Z}{{\mathbb Z}}

\newcommand{\ii}{\mathrm{i}}
\newcommand{\ee}{\mathrm{e}}

\newcommand{\vev}[1]{\langle #1 \rangle}

\newcommand{\diag}{\mathrm{diag}\,}

\newcommand{\cF}{{\mathcal{F}}}

\newcommand{\cN}{{\mathcal{N}}}
\newcommand{\cO}{{\mathcal{O}}}

\newcommand{\cS}{{\mathcal{S}}}
\newcommand{\cT}{{\mathcal{T}}}

\newcommand{\cW}{{\mathcal{W}}}
\newcommand{\cY}{{\mathcal{Y}}}

\newcommand{\one}{{\rm 1\kern -.9mm l}}

\newcommand{\eud}{\epsilon_1\epsilon_2}

\newdimen\tableauside\tableauside=1.0ex
\newdimen\tableaurule\tableaurule=0.4pt
\newdimen\tableaustep
\def\phantomhrule#1{\hbox{\vbox to0pt{\hrule height\tableaurule
width#1\vss}}}
\def\phantomvrule#1{\vbox{\hbox to0pt{\vrule width\tableaurule
height#1\hss}}}
\def\sqr{\vbox{%
  \phantomhrule\tableaustep
\hbox{\phantomvrule\tableaustep\kern\tableaustep\phantomvrule\tableaustep}%
  \hbox{\vbox{\phantomhrule\tableauside}\kern-\tableaurule}}}
\def\squares#1{\hbox{\count0=#1\noindent\loop\sqr
  \advance\count0 by-1 \ifnum\count0>0\repeat}}
\def\tableau#1{\vcenter{\offinterlineskip
  \tableaustep=\tableauside\advance\tableaustep by-\tableaurule
  \kern\normallineskip\hbox
    {\kern\normallineskip\vbox
      {\gettableau#1 0 }%
     \kern\normallineskip\kern\tableaurule}%
  \kern\normallineskip\kern\tableaurule}}
\def\gettableau#1 {\ifnum#1=0\let\next=\null\else
  \squares{#1}\let\next=\gettableau\fi\next}
\newcommand{\eu}{\epsilon_1}
\newcommand{\ed}{\epsilon_2}

\title{Modular anomaly equation, heat kernel and S-duality in $\cN=2$ theories
}
\author{M. Bill\'o$^1$, M. Frau$^{1}$, L. Gallot$^{2}$, A. Lerda$^{3}$, I. Pesando$^{1}$
\\
\vskip 0.2cm
$^1$ Universit\`a di Torino, Dipartimento di Fisica
and I.N.F.N. - sezione di Torino \\
Via P. Giuria 1, I-10125 Torino, Italy\\
\vskip 0.2cm
$^2$LAPTH, Universit\'e de Savoie, CNRS\\
9, Chemin de Bellevue,
74941 Annecy le Vieux Cedex, France\\
\vskip 0.2cm
$^3$Universit\`a del Piemonte Orientale, Dipartimento di Scienze e Innovazione Tecnologica, \\
and I.N.F.N. - Gruppo Collegato di Alessandria - sezione di Torino\\
Viale T. Michel  11, I-15121 Alessandria, Italy\\
\vspace{0.25cm}
\email{billo,frau,lerda,ipesando@to.infn.it; laurent.gallot@lapth.cnrs.fr} 
}

\abstract{We investigate $\epsilon$-deformed $\cN=2$
superconformal gauge theories in four dimensions, focusing on the $\cN=2^*$ and $N_f=4$ SU(2) cases. 
We show how the modular anomaly equation obeyed by the deformed prepotential can be efficiently
used to derive its non-perturbative expression starting from the perturbative one. 
We also show that the modular anomaly equation implies that S-duality is implemented by means of an exact Fourier transform even for arbitrary values of the deformation parameters, 
and then we argue that it is possible, perturbatively in the deformation, to choose appropriate variables such that it reduces to a Legendre transform.}

\preprint{LAPTH-040/13}

\begin{document}

\section{Introduction}
\label{secn:intro}
Gauge theories with rigid $\cN=2$ supersymmetry in four dimensions represent one of the main
areas where substantial progress toward a non-perturbative description has been made, 
starting with the seminal papers of Seiberg and Witten (SW) \cite{Seiberg:1994rs,Seiberg:1994aj}. 
In recent years much attention has been devoted to the deformation of these theories by means of 
the so-called $\epsilon$ (or $\Omega$) background. This anti-symmetric tensor background breaks Lorentz 
invariance and can be generically characterized by two parameters $\epsilon_1$ and $\epsilon_2$. 
Originally, it was used by Nekrasov
\cite{Nekrasov:2002qd}\nocite{Nekrasov:2003af,Nekrasov:2003rj}\,-\,\cite{Losev:2003py} as a regulator in the explicit computations, by means of localization techniques, of the
multi-instanton contributions to the partition function%
\footnote{Here $a$ stands or the vacuum expectation values of the adjoint complex scalar in 
the $\cN=2$ gauge multiplet along the Cartan directions, which parametrize the moduli space in the Coulomb branch; in this paper we will be concerned with rank one cases, so that we will have a single $a$. In presence of matter multiplets, the expressions depend on the masses 
of the latter as well. They will also depend on the dynamically generated scale $\Lambda$ or 
(in the conformal cases) on the bare coupling $\tau_0$.}
$Z(a;\epsilon)$ and to the prepotential of the low-energy effective theory
\begin{equation}
\label{Fdef}
F(a;\epsilon) = F_{\mathrm{cl}}(a) - \eud\, \log Z(a;\epsilon)
\end{equation}
where $F_{\mathrm{cl}}$ is the classical term.
The prepotential can be expanded in powers of $\epsilon$, and it is convenient to organize such an expansion as follows:
\begin{equation}
\label{Fexp}
F(a;\epsilon) = F_{\mathrm{cl}}(a) + \sum_{n,g=0}^\infty 
(\epsilon_1+\epsilon_2)^{2n}\,(\eud)^g\,F^{(n,g)}(a)~,
\end{equation}
where the coefficients $F^{(n,g)}$ account for the
perturbative and non-perturbative contributions. The usual prepotential 
$\cF$ of the SW theory is obtained by setting the $\epsilon$ regulators to zero, 
that is $\cF(a) \equiv F_{\mathrm{cl}}(a)+ F^{(0,0)}(a)$ in the above notation. 
In this way one can derive the explicit instanton expansion 
and successfully compare it with the expression obtained 
by geometrizing the monodromy and the duality features of the low-energy theory in terms of a SW curve.
In the latter treatment, the vacuum expectation values $a$ and their S-duals%
\footnote{Throughout this paper, we will denote the S-dual of a quantity $X$ as $\cS[X]$ or, whenever typographically clear, as $\widetilde X$.} 
$\tilde a$ arise as periods of a suitable meromorphic differential $\lambda$ along a symplectic basis $(A,B)$ of one-cycles of the SW curve, and the prepotential $\cF$ is such that
\begin{equation}
\label{Ftoaat}
2\pi\ii\,\tilde a = \frac{\partial \cF}{\partial a}~.
\end{equation}
The duality group of the effective theory is contained in the 
$\mathrm{Sp}(2r,\Z)$ redefinitions of the symplectic basis, that reduce to $\mathrm{Sl}(2,\Z)$ in the rank $r=1$ case. 
The $\cS$ generator of this group exchanges $a$ and $\tilde a$, and the S-dual description of the theory is given through the Legendre transform of the prepotential:
\begin{equation}
\label{LTFsw}
{\widetilde \cF}(\tilde a)= \cF(a) - 2 \pi\ii \tilde a\, a~,
\end{equation}  
where in the right hand side $a$ has to be expressed in terms of $\tilde a$ by inverting 
(\ref{Ftoaat}).

It is natural to wonder about the physical meaning of the prepotential $F(a;\epsilon)$ at 
finite values of the $\epsilon$-deformation.  
This question was first addressed in the particular case $\epsilon_2 = -\epsilon_1$, when only the 
$n=0$ terms in (\ref{Fexp}) contribute. In \cite{Nekrasov:2002qd}\nocite{Nekrasov:2003rj}\,-\,\cite{Losev:2003py} 
it was proposed that they correspond to gravitational F-terms in the effective action 
of the form $F^{(0,g)}\,\cW^{2g}$, where $\cW$ is the chiral Weyl superfield containing the 
graviphoton field strength as its lowest component \cite{Antoniadis:1993ze}.
This statement can be understood by realizing the $\cN=2$ gauge theories 
on the world-volume of stacks of D-branes in backgrounds with 
four orbifolded internal directions.
In such a microscopic string set-up it is possible to compute 
the non-perturbative corrections provided by D-instantons \cite{Billo:2002hm}
and show that the $\epsilon$-background corresponds to a Ramond-Ramond three-form which appears in 
the effective theory as the graviphoton \cite{Billo:2006jm}; the above gravitational F-terms are thus directly accounted for%
\footnote{For another string theory interpretation of the $\Omega$ background, see
\cite{Hellerman:2011mv,Hellerman:2012zf}.}.

On the other hand, the low-energy $\cN=2$ theories can be  
``geometrically engineered'' in Type II string theory \cite{Kachru:1995fv}\nocite{Klemm:1996bj}\,-\,\cite{Katz:1996fh}
in terms of closed strings on suitable non-compact ``local'' Calabi-Yau (CY) manifolds 
whose complex structure moduli $t$ encode the gauge-invariant moduli of 
the SW theory (and are thus non-trivially related to the Coulomb branch parameters $a$). 
In such CY compactifications, the graviphoton F-terms of the four-dimensional effective action 
are captured by {topological} string amplitudes 
as $F_{\mathrm{top}}^{(g)}(t)\,\cW^{2g}$ \cite{Antoniadis:1993ze,Bershadsky:1993cx}. 
For local CY realizations of $\cN=2$ gauge theories, the coefficients $F^{(0,g)}(a)$ 
are therefore identified with the topological amplitudes $F_{\mathrm{top}}^{(g)}(t)$ and the expansion 
of the deformed prepotential $F(a;\epsilon)$ in powers of $\eud$ corresponds to the genus expansion 
of the topological string vacuum amplitude with a string coupling constant
\begin{equation}
\label{gs}
g_s^2=\eud~.
\end{equation}

The link to the topological string brings an important bonus: the amplitudes $F_{\mathrm{top}}^{(g)}$  
satisfy a ``holomorphic anomaly equation'' \cite{Bershadsky:1993ta,Bershadsky:1993cx}. 
In fact, despite their apparent holomorphicity in $t$, 
they also develop a dependence on $\bar t$ due to contributions from the boundary of the genus 
$g$ moduli space, and thus should be more properly denoted as $F_{\mathrm{top}}^{(g)}(t,\bar t)$. 
The holomorphic anomaly equation, which relates amplitudes at different genera, 
can be written \cite{Witten:1993ed} as a linear equation, similar in structure to the heat equation, for the topological partition function
\begin{equation}
\label{topwf}
Z_{\mathrm{top}}(t,\bar t; g_s) = 
\exp\Bigl(
-\sum_{g=0}^\infty g_s^{2g-2}\,F_{\mathrm{top}}^{(g)}(t,\bar t)
\Bigr)~,
\end{equation}
which in the local CY case can be connected to the deformed partition function 
$Z(a;\epsilon)$ of the corresponding $\cN=2$ theory. 

The holomorphic anomaly can be understood 
by assuming that the moduli space of complex structures has to be quantized and
that the partition function behaves as a wave function,
with $g_s^2$ playing the r\^ole of $\hbar$ \cite{Witten:1993ed}\nocite{Aganagic:2006wq}\,-\,\cite{Gunaydin:2006bz}. 
This assumption requires that the partition function transforms consistently 
under canonical transformations acting on the moduli phase space.
Parametrizing the phase space by the periods $a$ and $\tilde a$, on which the duality symmetries 
act as symplectic transformations, one can thus determine the effect of the latter 
on $Z(a;\epsilon)$. In particular, the S-duality exchanging $a$ and $\tilde a$ should be represented 
on $Z(a;\epsilon)$ as a Fourier transform:
\begin{equation}
\label{FTZ}
\widetilde Z(\tilde a;\epsilon)\simeq \int \!dx~\ee^{\frac{2\pi\ii\tilde a\, x}{g_s^2\phantom{|}}}
\, Z(x;\epsilon)~.
\end{equation}
Using (\ref{Fdef}) and (\ref{Fexp}), one can evaluate the integral in the saddle-point approximation 
for small values of $g_s$ and show that only at the leading order the S-duality reduces to a 
Legendre transform as in (\ref{LTFsw}).
By comparing the description based on $(a,\tilde a)$ with the
one employing the global cooordinates $(t,\bar t)$, it is possible to relate the non-holomorphicity 
of the amplitudes $F_{\mathrm{top}}^{(g)}(t,\bar t)$ to the failure of modularity for the $F^{(0,g)}(a)$ 
and to argue that the amplitudes should be given either in terms of modular but almost holomorphic expressions (in the global coordinates description) or in terms of holomorphic but quasi-modular 
forms (in the period basis) \cite{Aganagic:2006wq,Gunaydin:2006bz}.

The general picture we have outlined above has been pursued and applied
to many $\cN=2$ models, ranging from SU(2) to higher rank pure 
gauge theories, as well as to theories with matter in various representations of the gauge group. 
It is of course a highly non-trivial task to \emph{explicitly} compute the deformed prepotential 
in a given model, and the various approaches can cooperate to this aim. On the microscopic 
side, efficient tools have been devised to implement the localization techniques and compute non-perturbative corrections \cite{Nakajima:2003uh}\nocite{Flume:2004rp,
Nekrasov:2004vw,Marino:2004cn,Billo:2009di,Fucito:2009rs,Billo':2010bd,Billo:2010mg}\,-\,\cite{Ghorbani:2010ks}. 
On the topological side, results can be obtained from a direct integration of the 
holomorphic anomaly equation, supplemented with appropriate boundary conditions \cite{Klemm:2002pa}\nocite{Huang:2006si,Grimm:2007tm, Huang:2009md}\,-\,\cite{Krefl:2010fm}. 

In the case of $\cN=2$ \emph{superconformal} gauge theories, 
an extremely interesting interpretation of the $\epsilon$-deformation, going
under the name of AGT relation, has been uncovered \cite{Alday:2009aq}\nocite{Gaiotto:2009we,Wyllard:2009hg,Mironov:2009qt}\,-\,\cite{Mironov:2009by}: 
the generalized prepotential $F(a;\epsilon)$ equals the logarithm of conformal 
blocks in a two-dimensional Liouville theory. 
The central charge of the Liouville 
theory and the dimensions of the operators are $\epsilon$-dependent and 
the number of inserted operators, as well as the genus of the two-dimensional surface on which the
conformal blocks are defined, depend on the gauge group and matter content.
In order to realize a generic central charge for the Liouville theory, 
it is necessary to consider the case with non-vanishing 
\begin{equation}
\label{sis}
s= \epsilon_1 + \epsilon_2~,
\end{equation}
which means that the full expansion (\ref{Fexp}) and not just its $n=0$ terms must be taken into account. The $F^{(n,g)}(a)$ coefficients with $n\not= 0$  correspond to so-called ``refined'' 
topological amplitudes $F^{(n,g)}_{\mathrm{top}}(t,\bar t)$. They
appear in F-terms of the form  
$F^{(n,g)}_{\mathrm{top}}(t,\bar t)\, \cY^{2n}\,\cW^{2g}$, involving a chiral 
superfield $\cY$ composed out
of extra vector multiplets \cite{Antoniadis:2010iq,Antoniadis:2013bja}, and obey a generalized 
holomorphic anomaly equation \cite{Huang:2006si}\nocite{Grimm:2007tm,Huang:2009md}\,-\,\cite{Krefl:2010fm}. 
The AGT relation maps the duality properties, and in particular the strong/weak-coupling S-duality, of the $\cN=2$ superconformal theories into the modular transformation properties of the Liouville conformal blocks; this makes  the study of the duality properties of these superconformal theories at generic values of the $\epsilon$-parameters even more interesting. 

The simplest canonical examples of four-dimensional superconformal gauge theories are given 
by SU(2) SYM theory with either $N_f=4$ adjoint matter hypermultiplets or with one adjoint hypermultiplet, the latter case being also known as the $\cN=2^*$ theory. 
These theories have vanishing $\beta$-function, but, when the hypermultiplets are massive, 
they receive both perturbative and non-perturbative corrections. For vanishing $\epsilon$-deformations, via the SW description, it is possible to obtain the prepotential $\cF$ as an exact function of 
the bare coupling $\tau_0$, and show that under S-duality, which on the bare coupling acts by $\tau_0\to -1/\tau_0$, the prepotential and its S-dual are 
related precisely by a Legendre transform, as in (\ref{LTFsw}) so that also the effective
coupling $\tau$ is mapped by $\cS$ into $-1/\tau$ \cite{Minahan:1997if,Billo':2011pr}.

Using  the various approaches described above, much progress has been made in obtaining exact expressions for the generalized prepotential terms $F^{(n,g)}(a,\epsilon)$ and in 
deriving their modular properties. In particular, using the topological string point of view,
the generalized holomorphic anomaly equation that applies to these superconformal cases
has been analyzed in \cite{Huang:2011qx}\nocite{Huang:2012kn}\,-\,\cite{Huang:2013eja}. This holomorphic anomaly translates into an anomalous modular behavior of the $F^{(n,g)}(a,\epsilon)$'s with respect 
to the bare coupling $\tau_0$ which can only occur through the second Eisenstein series 
$E_2(\tau_0)$. The dependence on $E_2$ is described by a modular anomaly equation of the form
\cite{Huang:2011qx}\nocite{Huang:2012kn,Huang:2013eja}\,-\,\cite{Billo:2013fi}
\begin{equation}
 \partial_{E_2}F^{(n,g)}= -
\frac{1}{24k} \sum_{n_1=0}^n\sum_{g_1=0}^g
\partial_aF^{(n_1,g_1)}\,\partial_a F^{(n-n_1,g-g_1)}+
\frac{1}{24k}\partial_a^2F^{(n,g-1)}~,
\label{maegen}
\end{equation}
with $k=2$ for the $\cN=2^*$ theory, and $k=1$ for the $N_f=4$ one. 
The last term in the right hand side of (\ref{maegen}) is absent when $g=0$. This case happens 
in the so-called Nekrasov-Shatashvili (NS) limit $\ed\to 0$ with $\eu$ finite 
\cite{Nekrasov:2009rc}, which selects precisely the coefficients $F^{(n,0)}$.
Via the modular anomaly equation, exact expressions in terms of modular forms 
have been obtained for the first few $F^{(n,g)}$ coefficients at generic values of the 
$\epsilon$ parameters in the massless cases \cite{Huang:2011qx,Huang:2012kn}, and for the massive 
$\cN=2^*$ theory in the NS limit \cite{Huang:2013eja}.
Explicit results have been obtained also from the microscopic point of view, computing the non-perturbative instanton corrections by means of localization techniques. In particular, in \cite{Billo:2013fi} 
the expression in terms of modular forms of the first few $F^{(n,g)}$'s, for generic deformation and masses, was inferred from their instanton expansion; 
these findings suggest a recursion relation among the coefficient of the expansion of the prepotential for large $a$ 
which is equivalent to the modular anomaly equation (\ref{maegen}). 
Finally, to obtain explicit expressions and uncover the modular properties of the deformed theories, one can exploit the AGT relation, as recently done in \cite{KashaniPoor:2012wb,Kashani-Poor:2013oza},
or the deformed matrix models \cite{Baek:2013oia}. 

In this paper we address various issues. 
In Section~\ref{secn:burger} we present an efficient method to obtain the 
exact expressions of the $F^{(n,g)}$'s based on the modular anomaly equation. We show that 
this equation is equivalent to the heat equation on the non-classical part of the partition function, 
and use the heat kernel to express the prepotential in terms of its ``boundary'' value obtained by 
disregarding all terms that involve the Eisenstein series $E_2$. With the knowledge of the 
boundary value obtained from the perturbative 1-loop result (and possibly the very first few 
instanton terms), one can derive the exact expression in terms of modular forms of the 
prepotential up to a very high order in the expansion for large $a$. 

The knowledge of the explicit form of the prepotential is crucial in analyzing its modularity 
properties, and in particular the way S-duality is implemented in the full deformed quantum 
theory. It is readily established that the generalized prepotential $F(a;\epsilon)$
and its S-dual $\widetilde F(\tilde a;\epsilon)$ are no longer related to each other by a Legendre transform when $g_s\not= 0$. Recently in \cite{Galakhov:2012gw} it
has been proposed that for generic values of the $\epsilon$-parameters the S-duality acts as 
a modified Fourier transform on the deformed partition function; however, more recently, in \cite{Nemkov:2013qma} it has been conjectured, also on the basis of the explicit results of \cite{Billo:2013fi}, that it is in fact exactly given by the exact Fourier transform (\ref{FTZ}), 
which would thus be valid also for $\epsilon_1+\epsilon_2\not = 0$ and for massive superconformal 
$\cN=2$ theories. 
In Section~\ref{sft} we show that the modular anomaly equation (\ref{maegen}) implies that 
this conjecture is indeed correct. 

In Section~\ref{spa} we take a further step and show that it is possible to introduce a 
modified prepotential $\widehat F(a;\epsilon)$, determined order by order in $g_s^2$, in such a 
way that the S-duality gets reformulated as a Legendre transform of this new quantity. 
This possibility had already been pointed out in \cite{Billo:2013fi}, where 
$\widehat F$ was determined by a rather involved procedure imposing order by order the consistency 
condition that $\cS^2[a]=-a$. 
Here this result follows much more naturally from the Fourier transform property of $Z(a;\epsilon)$. 
We find this observation quite intriguing, even if we do not yet have a clear physical 
interpretation of $\widehat F$, and in Section~\ref{spa} and in the Conclusions we further comment 
on it. Finally, we collect 
in the Appendices several technical details that are useful for the explicit calculations.

\section{Solution of the modular anomaly equation and heat kernel}
\label{secn:burger}
In this section we discuss how to solve the modular anomaly equation (\ref{maegen}) in two significant 
models: the mass-deformed $\cN=4$ SU(2) SYM theory, also known as $\cN=2^*$ theory, 
and the $\cN=2$ SU(2) SYM theory with $N_f=4$ fundamental flavors. In both cases we show 
how to reconstruct the generalized effective prepotential, 
including its non-perturbative terms, from the perturbative ones%
\footnote{A related technique was illustrated in \cite{Minahan:1997if} for the $\cN=2^*$ theory at vanishing $\epsilon$; a similar question can be addressed also in the matrix model approach
\cite{Ferrari:2012qj,Bourgine:2012bv}.}.

\subsection{The $\cN=2^*$ SU(2) theory}
\label{subsecn:n2star}
The $\cN=2^*$ SYM theory with gauge group SU(2) describes 
the interactions of an $\cN=2$ gauge vector multiplet with a massive $\cN=2$ hypermultiplet 
in the adjoint representation of SU(2). After giving a vacuum expectation value to the
scalar field $\phi$ of the vector multiplet:
\begin{equation}
 \vev{\phi} = \diag(a,-a)~,
\label{vev}
\end{equation}
and introducing the parameters $\epsilon_1$ and $\epsilon_2$ of the Nekrasov
background \cite{Nekrasov:2002qd}\nocite{Nekrasov:2003af}\,-\,\cite{Nekrasov:2003rj}, the deformed prepotential $F$
takes the form (\ref{Fexp}) where
\begin{equation}
 F_{\mathrm{cl}}= 2\pi\ii\tau_0\,a^2~,
\label{Fcl}
\end{equation}
$\tau_0$ being the bare gauge coupling constant.
The coefficients $F^{(n,g)}$ of the $\epsilon$-expansion account for the perturbative and non-perturbative contributions, and are functions of $a$, of the hypermultiplet mass $m$ and 
of $\tau_0$ through the Eisenstein series $E_2$, $E_4$
and $E_6$.%
\footnote{For these modular functions and for the Jacobi $\theta$-functions appearing in the next
subsection, we use the conventions given in Appendix~A of \cite{Billo:2013fi}.} As shown in \cite{Billo:2013fi,Huang:2013eja}, they satisfy the modular anomaly 
equation (\ref{maegen}) with $k=2$.
Defining
\begin{equation}
\begin{aligned}
 \varphi_0 &\equiv F_{\mathrm{cl}}-F=- \sum_{n,g=0}^\infty 
(\epsilon_1+\epsilon_2)^{2n}\,(\epsilon_1\epsilon_2)^g\,F^{(n,g)}~,
\end{aligned}
\label{phi0}
\end{equation}
it is immediate to show that (\ref{maegen}) becomes
\begin{equation}
 \partial_{E_2} \varphi_0 =\frac{1}{48} (\partial_a\varphi_0)^2
+\frac{\epsilon_1\epsilon_2}{48}\,\partial_a^2\varphi_0
\label{Dphi0}
\end{equation}
which is the homogeneous Kardar-Parisi-Zhang (KPZ) equation in one space dimension \cite{Kardar:1986xt}.
To write it in the standard form, namely
\begin{equation}
 \partial_t \varphi_0 = \frac12 \big(\partial_x \varphi_0\big)^2 +\nu\,\partial_x^2\varphi_0~,
\label{KPZ}
\end{equation}
it is enough to set
\begin{equation}
 t=\frac{E_2}{24}~,~~~x=a~~~\,\mbox{and}~~~\,\nu=\frac{\eud}{2}~.
\label{settings}
\end{equation}
By taking a further derivative of (\ref{KPZ}) with respect to $x$, one finds the viscous Burgers equation 
\cite{Burgers}
\begin{equation}
 \partial_t u + u\,\partial_x u = \nu\,\partial_x^2u
\label{burgers}
\end{equation}
where $u\equiv-\partial_x \varphi_0$. It is well-known that the non-linear
Burgers equation (\ref{burgers}) can be mapped 
into a linear parabolic equation by means of the Hopf-Cole transformation \cite{Hopf,Cole}.
The same is true also for the non-linear KPZ equation (\ref{KPZ}). Indeed, writing
\begin{equation}
 \varphi_0 = \eud\,\log \Psi~,
\label{Psi}
\end{equation}
one easily obtains 
\begin{equation}
 \partial_t\Psi - \frac{\eud}{2}\, \partial_a^2 \Psi=0
\label{heat}
\end{equation}
which is the linear heat conduction equation.
Note that, using (\ref{phi0}), we can rewrite (\ref{Psi}) as
\begin{equation}
 \Psi = \exp\Big(\frac{\varphi_0}{\eud}\Big)= \exp\Big(\!\!-\frac{F-F_{\mathrm{cl}}}{\eud}\Big)~,
\label{Psi1}
\end{equation}
{from} which we read that $\Psi$ is the non-classical part of the generalized partition 
function of the theory. The fact that the partition function is related to a solution of the 
heat equation was already noticed in \cite{Witten:1993ed}, even if in a different context, and 
more recently also in \cite{Huang:2011qx}.

The general solution of the heat equation (\ref{heat}) 
can be written as the convolution of the heat kernel
\begin{equation}
G(x;t)= \frac{1}{\sqrt{2\pi\eud t\phantom{|}}}\,\exp\Big(\!\!-\frac{x^2}{2\eud t}\Big)~,
\label{heatkernel}
\end{equation}
with an ``initial'' condition $\Psi_0\equiv \Psi|_{t=0}$, namely 
as
\begin{equation}
\Psi(a;t) =\big(G * \Psi_0\big)(a;t)
\label{convolution}
\end{equation}
or, more explicitly, as
\begin{equation}
\exp\Big(\frac{\varphi_0(a;t)}{\eud}\Big) =\frac{1}{\sqrt{2\pi\eud t\phantom{|}}}\,\int_{-\infty}^{+\infty}\!\!dy~\exp\Big(\!
{-\frac{(a-y)^2}{2\eud t}+\frac{\varphi_0(y;0)}{\eud}}\Big)~.
\label{sol}
\end{equation}
This formula allows us to reconstruct the dependence on $E_2$ of the non-classical part of the 
generalized prepotential starting from $\varphi_0$
evaluated at $E_2=0$. Let us now give some details.

As discussed in \cite{Billo:2013fi} (see also \cite{Huang:2011qx}) the perturbative part of the
generalized prepotential is
 \begin{equation}
\begin{aligned}
F_{\mathrm{pert}} &\equiv- \varphi_0\big|_{\mathrm{pert}} = \epsilon_1\epsilon_2\Big[
\gamma_{\epsilon_1,\epsilon_2}(2a)+\gamma_{\epsilon_1,\epsilon_2}(-2a)
-\gamma_{\epsilon_1,\epsilon_2}(2a+\widetilde m) - \gamma_{\epsilon_1,\epsilon_2}(-2a+\widetilde m)
\Big] 
\end{aligned}
 \label{F1loop1}
\end{equation}
where 
\begin{equation}
 \widetilde m = m+\frac{\epsilon_1+\epsilon_2}{2}
\label{tildem}
\end{equation}
is the equivariant mass parameter \cite{Okuda:2010ke}, and
\begin{equation}
 \gamma_{\epsilon_1,\epsilon_2}(x) = \left.\frac{d}{ds}\Big(\frac{\Lambda^s}{\Gamma(s)}
\int_0^\infty \frac{dy}{y}\frac{y^s\,\ee^{-yx}}{(\ee^{-\epsilon_1 y}-1)
(\ee^{-\epsilon_2 y}-1)}\Big)\right|_{s=0}
\label{gammae1e2}
\end{equation}
is related to the logarithm of the Barnes double $\Gamma$-function \cite{Nekrasov:2003rj,Nakajima:2003uh,Huang:2011qx}. Choosing a branch for this logarithm and
expanding for small values of $\epsilon_1$ and $\epsilon_2$, one finds \cite{Billo:2013fi}
\begin{equation}
\varphi_0\big|_{\mathrm{pert}} = -\frac{1}{2}\,h_0 \log\frac{4a^2}{\Lambda^2}+
\sum_{\ell=1}^\infty \frac{~h_\ell^{(0)}}{2^{\ell+1}\,\ell}\,\frac{1}{a^{2\ell}} 
\label{phi0p}
\end{equation}
where the first few coefficients are
\begin{align}
h_0\phantom{^{0}\,} &=\frac{1}{4}\big(4m^2-s^2\big)\phantom{\Big|}~,
\label{h00}\\
h_1^{(0)} &
=\frac{1}{12}h_0\big(h_0+\eud\big)~,
\label{h10}\\
h_2^{(0)} &
=\frac{1}{120}h_0\big(h_0+\eud\big)\big(2h_0-s^2+3\eud\big)
\phantom{\Big|}~,\label{h20}\\
h_3^{(0)} &
=\frac{1}{672}h_0\big(h_0+\eud\big)\big(3h_0^2-4h_0s^2+11 h_0 \eud+
10(\eud)^2-10s^2\eud+2s^4\big)\phantom{\Big|}~,\label{h30}
\end{align}
with $s=\eu+\ed$. In Appendix \ref{app:a} we also give the expression for $h_4^{(0)}$.

Among the various terms in (\ref{phi0p}), the logarithmic one
plays a distinguished r\^ole because it is exact at 1-loop and does not receive any 
non-perturbative correction%
\footnote{This is the reason why the coefficient $h_0$ does not carry the superscript
$^{(0)}$. On the contrary the terms proportional to $a^{-2\ell}$
get corrected by instantons and their exact coefficients $h_\ell$'s will be the non-perturbative
completion of the $h_\ell^{(0)}$'s given in (\ref{h10})-(\ref{h30}).}. 
Therefore, it is natural to expect that if we take it as the ``initial condition'' at $t=0$,
we can generate all structures of the prepotential that only depend
on $E_2$. Indeed, if in (\ref{sol}) we take
\begin{equation}
\varphi_0(y;0) 
\simeq -\frac{1}{2}\,h_0 \log\frac{4y^2}{\Lambda^2}~,
\label{phi00}
\end{equation}
then
\begin{equation}
 \exp\Big(\frac{\varphi_0(a;t)}{\eud}\Big) \simeq
 \frac{1}{\sqrt{2\pi\eud t\phantom{|}}}\,\int_{-\infty}^{+\infty}\!\!dy~
 \exp\Big(\!-\frac{(a-y)^2}{2\eud t}\Big)
~\Big(\frac{2y}{\Lambda}\Big)^{-\frac{h_0}{\eud}}~.
\label{sol0} 
\end{equation}
With a suitable change of variables, we can recognize in (\ref{sol0}) 
the integral representation of the parabolic cylinder functions (see Appendix \ref{app:b}).
Carrying out the integration over $y$, we can organize the result as an expansion in
inverse powers of $a$ as follows
\begin{equation}
 \exp\Big(\frac{\varphi_0(a;t)}{\eud}\Big) \simeq \Big(\frac{2a}{\Lambda}\Big)^q
\sum_{\ell=0}^\infty \frac{q(q-1)\cdots(q-2\ell+1)}{2^\ell\,\ell!}\,\frac{(\eud t)^\ell}{a^{2\ell}}
\label{sol1} 
\end{equation}
where for convenience we have defined
\begin{equation}
 q=-\frac{h_0}{\eud}~.
\label{q}
\end{equation}
Taking the logarithm of (\ref{sol1}) and reinstating $E_2$ according to (\ref{settings}),
after simple algebra we have
\begin{equation}
\begin{aligned}
\varphi_0 &\simeq -\frac{1}{2}\,h_0 \log\frac{4a^2}{\Lambda^2}+
\sum_{\ell=1}^\infty \frac{h_\ell}{2^{\ell+1}\,\ell}\,\frac{1}{a^{2\ell}} 
\end{aligned}
\label{phi01}
\end{equation}
where the first few coefficients are
\footnote{It is worth noting that, while the coefficients of $a^{-2\ell}$ in the expansion of 
$\exp\Big(\frac{\varphi_0(a;t)}{\eud}\Big)$ given in (\ref{sol1}) are polynomials of degree 
$2\ell$ in $h_0$, the coefficients $h_\ell$ in the expansion of $\varphi_0$ are polynomials
of degree $\ell+1$ in $h_0$. This is simply due to dimensional reasons. 
The cancellations that occur in passing from $\exp\Big(\frac{\varphi_0(a;t)}{\eud}\Big)$ to 
$\varphi_0$ are a consequence of the properties of the 
parabolic cylinder functions as explained in Appendix~\ref{app:b}.}
 \begin{align}
 h_1 &\simeq \frac{1}{12}h_0\big(h_0+\eud\big)\,E_2~,\label{h11}\\
 h_2 &\simeq \frac{1}{144}h_0\big(h_0+\eud\big)\big(2h_0+3\eud\big)\,E_2^2~,\label{h21}\\
 h_3 &\simeq \frac{1}{1728}h_0\big(h_0+\eud\big)\big(5h_0^2+17h_0\,\eud+15(\eud)^2\big)\,E_2^3~,
\label{h31}
 \end{align}
and so on and so forth. As expected, the $h_\ell$'s turn out to be quasi-modular forms of weight $2\ell$ made only of powers of $E_2$. Since in this model
there is only one form of weight 2, namely $E_2$ itself, the above expression for $h_1$ must be exact. 
This observation is further confirmed by the fact that in the perturbative limit $E_2\to 1$ 
(\ref{h11}) correctly reduces to $h_1^{(0)}$ given in (\ref{h10}). Indeed
\begin{equation}
\Theta_1\equiv h_1^{(0)} - h_1\big|_{E_2\to 1} = 0
\label{Theta1}
\end{equation}
and in (\ref{h11}) we can replace $\simeq$ with $=$. 

On the contrary, the $h_\ell$'s with $\ell\geq 2$ given above are not exact, since there are
other modular forms of weight $2\ell$ beside $E_2^{\ell}$ that could or should be present. 
For instance, for $\ell=2$ we have also the Eisenstein series $E_4$. 
Moreover, the perturbative limit of (\ref{h21}) does not coincide
with $h_2^{(0)}$ given in (\ref{h20}):
\begin{equation}
\Theta_2\equiv h_2^{(0)} - h_2\big|_{E_2\to 1} = 
\frac{1}{720}h_0\big(h_0+\eud\big)\big(2h_0+3\eud-6s^2\big)~.
\label{Theta2}
\end{equation}
To take this fact into account, we have to change the initial condition
in our heat-kernel formula (\ref{sol}) and use
\begin{equation}
 \varphi_0(y;0) 
\simeq -\frac{1}{2}\,h_0 \log\frac{4y^2}{\Lambda^2} + \frac{\Theta_2\,E_4}{16\,y^4}
\label{phi000}
\end{equation}
instead of (\ref{phi00}). If we do this, after simple algebra we obtain again (\ref{phi01}) with $h_1$ as in (\ref{h11}) but with $h_2$ and $h_3$ replaced by
\begin{align}
 h_2 &\simeq \frac{1}{144}h_0\big(h_0+\eud\big)\Big[\big(2h_0+3\eud\big)\,E_2^2
+\frac{1}{5}\big(2h_0+3\eud-6s^2\big)\,E_4\Big]~,\label{h22}\\
 h_3 &\simeq \frac{1}{1728}h_0\big(h_0+\eud\big)\Big[\big(5h_0^2+17h_0\,\eud+15(\eud)^2\big)\,E_2^3
\notag\\
&~~~~~~\qquad\qquad\qquad\qquad+\frac{3}{5}\big(2h_0+5\eud\big)\big(2h_0+3\eud-6s^2\big)\,E_2E_4\Big]~.
\label{h32}
 \end{align}
In this way we generate all terms depending on $E_2$ and those that are linear in $E_4$.
Actually, since there are no other possible forms of weight 4 other than $E_2^2$ and $E_4$, the above expression for $h_2$ is exact and its perturbative limit indeed coincides with $h_2^{(0)}$ given
in (\ref{h20}). Thus, in (\ref{h22}) we can substitute $\simeq$ with $=$.

Instead, the expression (\ref{h32}) for $h_3$ is still incomplete since there is a further modular
structure of weight 6 that has not yet appeared, namely $E_6$. Moreover, the perturbative 
limit of (\ref{h32}) does not reproduce the 1-loop result:
\begin{eqnarray}
 \Theta_3& \equiv& h_3^{(0)} - h_3\big|_{E_2,E_4\to 1} \label{Theta3}\\
&= &
\frac{1}{60480}h_0\big(h_0+\eud\big)\big(11h_0^2+59h_0\eud+60(\eud)^2-108h_0s^2
-270s^2\eud+180s^4\big)~.
\nonumber
\end{eqnarray}
This fact forces us to change once again the initial condition for the heat kernel and use
in (\ref{sol})
\begin{equation}
 \varphi_0(y;0) 
\simeq -\frac{1}{2}\,h_0 \log\frac{4y^2}{\Lambda^2} + \frac{\Theta_2\,E_4}{16\,y^4}
+\frac{\Theta_3\,E_6}{48\,y^6}
\label{phi0000}
\end{equation}
instead of (\ref{phi000}). By doing this, the resulting expression for $\varphi_0$
has the form (\ref{phi01}) with $h_1$ and $h_2$ as in (\ref{h11}) and (\ref{h22})
respectively, and with $h_3$ given by
\begin{equation}
 \begin{aligned}
  h_3&=\frac{1}{1728}\,h_0\big(h_0+\eud\big)\Big[\big(5h_0^2+17h_0\,\eud+15(\eud)^2\big)\,E_2^3
\\
&~~~~~~~~+\frac{3}{5}\big(2h_0+5\eud\big)\big(2h_0+3\eud-6s^2\big)\,E_2E_4\\
&~~~~~~~~+\frac{1}{35}\big(11h_0^2+59h_0\eud+60(\eud)^2-108h_0s^2
-270s^2\eud+180s^4\big)E_6\Big]~.
 \end{aligned}
\label{h3exact}
\end{equation}
Using (\ref{h00}) to express $h_0$ in terms of the hypermultiplet mass, one can check that our
results perfectly match those we obtained in \cite{Billo:2013fi} from explicit multi-instanton
calculations combined with the requirement of quasi-modularity.

This procedure can be further iterated with no conceptual difficulties to obtain the exact 
expressions of the higher coefficients $h_\ell$. In Appendix \ref{app:a} we provide some
details for the calculation of $h_4$. 
Of course, the algebraic complexity increases with $\ell$ but still
this method remains computationally very efficient. In particular we would like to remark that
the knowledge of the 1-loop prepotential (\ref{F1loop1}) together with the heat-kernel formula
(\ref{sol}) allows one to obtain the full non-perturbative expressions of $h_\ell$ up to
$\ell=5$. In $h_6$, which is a quasi-modular form of weight 12, there are
two independent structures of weight 12 not involving $E_2$, namely $E_6^2$ and $E_4^3$, and
thus the perturbative information is not sufficient to fix the relative coefficients. However,
combining this with the 1-instanton corrections \cite{Billo:2013fi}, one is able to resolve the
ambiguity and find $h_6$. In the same way one obtains all coefficients up to $h_{11}$; in order
to find $h_{12}$, which contains three different structures of weight 24 independent of $E_2$, namely
$E_6^4$, $E_6^2E_4^3$ and $E_4^6$, the 2-instanton results become necessary. This structure keeps repeating itself. 

Thus, we may conclude that the 1-loop and the first instanton corrections 
combined with the heat-kernel equation permit to reconstruct the exact generalized prepotential 
of the theory to a very high degree of accuracy in a systematic and algebraic fashion, generalizing
the method and the results of \cite{Minahan:1997if} to the case of arbitrary values of $\eu$ and $\ed$. 

\subsection{The SU(2) theory with $N_f=4$}
\label{subsecn:nf4}
We now repeat this analysis in the $\cN=2$ SU(2) SYM theory with four fundamental flavors. 
As is well-known, this
model has a vanishing 1-loop $\beta$-function and its conformal invariance is 
broken only by the flavor masses $m_f$ ($f=1,...,4)$. After giving a vacuum 
expectation value to the adjoint scalar field as in (\ref{vev}), the deformed
prepotential takes the form (\ref{Fexp}) with 
\begin{equation}
 F_{\mathrm{cl}}=\pi\ii\tau_0\,a^2~.
\label{FclNf4}
\end{equation}
Here we have used the standard normalization \cite{Seiberg:1994aj} for the classical term
(which differs by a factor of 2 from the $\cN=2^*$ one). 
The coefficients $F^{(n,g)}$ of the $\epsilon$-expansion 
depend on $a$, on the bare coupling $\tau_0$ through the Eisenstein series $E_2$,
$E_4$, $E_6$ and the Jacobi $\theta$-functions, and also on the hypermultiplet masses through the
SO(8) flavor invariants
\begin{equation}
 \label{invdef}
 \begin{aligned}
 R & = \frac 12 \sum_f m_f^2
~,\\
  T_1 & = \frac{1}{12} \sum_{f<f'} m_f^2 m_{f'}^2 - \frac{1}{24} \sum_f m_f^4
~,\\
 T_2 & = -\frac{1}{24} \sum_{f<f'} m_f^2 m_{f'}^2 + \frac{1}{48} \sum_f m_f^4 
-\frac 12 \prod_{f} m_f
~,\\
N & = \frac{3}{16} \sum_{f<f'<f''} m_f^2 m_{f'}^2 m_{f''}^2 - \frac{1}{96} 
 \sum_{f\not= f'} m_f^2 m_{f'}^4 + \frac{1}{96} \sum_f m_f^6~.
\end{aligned} 
\end{equation}
As shown in \cite{Billo:2013fi} (see also \cite{Huang:2011qx}), the coefficients
$F^{(n,g)}$ satisfy the modular anomaly equation (\ref{maegen}) with $k=1$.
We can therefore follow the same steps described in Section~\ref{subsecn:n2star} and prove that
\begin{equation}
 \exp\Big(\frac{\varphi_0(a;t)}{\eud}\Big) = 
\frac{1}{\sqrt{2\pi\eud t\phantom{|}}}\,\int_{-\infty}^{+\infty}\!\!dy~\exp\Big(\!
{-\frac{(a-y)^2}{2\eud t}+\frac{\varphi_0(y;0)}{\eud}}\Big)~.
\label{solnf4}
\end{equation}
This is the same as the $\cN=2^*$ equation (\ref{sol}), but with
\begin{equation}
 t= \frac{E_2}{12}
\label{tnf4}
\end{equation}
instead of $t=E_2/24$. 

As before, we use this equation to reconstruct $\varphi_0(a;t)$ starting from 
an ``initial'' condition at $t=0$. The latter can be taken from
the $N_f=4$ perturbative prepotential
\begin{equation}
 \varphi_0\big|_{\mathrm{pert}}=-\eud\Big[\gamma_{\eu\ed}(2a)+\gamma_{\eu\ed}(-2a)
-\sum_{f=1}^4\Big(\gamma_{\eu\ed}(a+\widetilde m_f)+\gamma_{\eu\ed}(-a+\widetilde m_f)\Big)\Big]
\label{f1loopnf4}
\end{equation}
where $\gamma_{\eu\ed}$ is given in (\ref{gammae1e2}) and 
the equivariant masses $\tilde m_f$ are defined as in (\ref{tildem}).
Expanding for large values  of $a$, we get
\begin{equation}
\varphi_0\big|_{\mathrm{pert}} = -\frac{1}{2}\,h_0 \log\frac{a^2}{\Lambda^2}+
\sum_{\ell=1}^\infty \frac{~h_\ell^{(0)}}{2^{\ell+1}\,\ell}\,\frac{1}{a^{2\ell}} 
\label{phi0pnf4}
\end{equation}
where the first few coefficients are \cite{Billo:2013fi}
\begin{align}
h_0\phantom{^{0}\,} & =\frac{1}{2}\big(4R - s^2 +\eud\big)~,\label{h0nf4}\\
h_1^{(0)} &=\frac{1}{6}h_0(h_0+\eud)-4T_1~,\label{h1nf4}\\
h_2^{(0)} &= \frac{1}{240}\big(16 h_0^3 +56 h_0^2 \eud -16 h_0^2 s^2 -960 h_0 T_1+28 h_0 (\eud)^2\notag\\
&~~~~~~~~~~~-16 h_0s^2\eud-1440 T_1\eud +480 T_1s^2-3s^2(\eud)^2+768 N 
\big)~.\label{h2nf4}
\end{align}
In Appendix~\ref{app:anf4} we also give the expression for $h_3^{(0)}$.
All coefficients, except $h_0$, receive non-perturbative corrections due to instantons which
can be explicitly computed using localization methods. 
Using as initial condition 
\begin{equation}
 \varphi_0(y;0) \simeq -\frac{1}{2}\,h_0 \log\frac{y^2}{\Lambda^2}~,
\label{initnf4}
\end{equation}
and following the same steps described in the $\cN=2^*$ theory, from (\ref{solnf4}) we 
generate all terms in the $h_\ell$'s which only depend on $E_2$. For example we find
\begin{align}
 h_1 &\simeq \frac{1}{6}h_0\big(h_0+\eud\big)\,E_2~,\label{h12nf4}\\
 h_2 &\simeq \frac{1}{36}h_0\big(h_0+\eud\big)\big(2h_0+3\eud\big)\,E_2^2~.\label{h22nf4}
\end{align}
In the perturbative limit, when $E_2\to 1$, these expressions do not reduce to (\ref{h1nf4})
and (\ref{h2nf4}), signaling the fact that other structures have to be considered in the
initial condition. Indeed, in the $N_f=4$ theory combining the usual modular properties
of the Jacobi $\theta$-functions with the triality transformations of the mass invariants
(\ref{invdef}), one can construct several modular forms. For
example, taking into account that the generators $\cT$ and $\cS$ of the modular group
act on $T_1$ and $T_2$ as \cite{Seiberg:1994aj} 
\begin{equation}
\begin{aligned}
& \cT~:~~ T_1~\to~T_1~,~~~\,T_2~\to~-T_1-T_2~,\\
&\cS~: ~~T_1~\to~T_2~,~~~~T_2~\to~T_1~,
\end{aligned}
\label{modular}
\end{equation}
and on the $\theta$-functions as
\begin{equation}
\begin{aligned}
& \cT~:~~ \theta_2^4~\to -\theta_2^4~,~~~~~\,\theta_3^4~\to~\theta_4^4~,~~~~~~~\,\theta_4^4~\to~\theta_3^4~,\\
&\cS~: ~~\theta_2^4~\to -\tau_0^2
\theta_2^4~,~~~\theta_3^4~\to-\tau_0^2\theta_3^4~,~~~\theta_4^4~\to-\tau_0^2\theta_4^4~,
\end{aligned}
\label{modulartheta}
\end{equation}
one can easily check that the combination
\begin{equation}
 T_1\,\theta_4^4-T_2\,\theta_2^4
\label{T1T2}
\end{equation}
is a modular form of weight 2. As such it could/should be present in the exact expression of $h_1$.
In the perturbative limit, $\theta_4\to 1$ and $\theta_2\to 0$, it simply reduces to $T_1$, and thus
by comparing with $h_1^{(0)}$ in (\ref{h1nf4}) we are led to change the initial 
condition (\ref{initnf4}) into
\begin{equation}
 \varphi_0(y;0) \simeq -\frac{1}{2}\,h_0 
\log\frac{y^2}{\Lambda^2}-\frac{T_1\,\theta_4^4-T_2\,\theta_2^4}{y^2}~.
\label{init1nf4}
\end{equation}
Inserting this into the heat-kernel formula (\ref{solnf4}) we obtain the final expression
for $h_1$:
\begin{equation}
 h_1 = \frac{1}{6}h_0\big(h_0+\eud\big)\,E_2-4\big(T_1\,\theta_4^4-T_2\,\theta_2^4\big)
~,\label{h1finnf4}
\end{equation}
and the following expression for $h_2$:
\begin{equation}
 h_2 \simeq \frac{1}{36}h_0\big(h_0+\eud\big)\big(2h_0+3\eud\big)\,E_2^2
-\frac{4}{3}\big(2h_0+3\eud\big)\big(T_1\,\theta_4^4-T_2\,\theta_2^4\big)\,E_2~.
\label{h23nf4}
\end{equation}
This fails to reproduce the perturbative limit (\ref{h2nf4}) since
\begin{equation}
 h_2^{(0)}-h_2\big|_{E_2\to1,\theta_4\to1,\theta_2\to0}=
\Theta_{2} + \Theta_{2}' T_1
\end{equation}
with
\begin{eqnarray}
\Theta_2&=&\frac{1}{720}\big(8h_0^3 +68 h_0^2 \eud -48 h_0^2 s^2 +24 h_0 (\eud)^2-48 h_0s^2\eud
-9s^2(\eud)^2+2304 N \big)~,\nonumber\\
\Theta_2'&=&-\frac{2}{3}\big(2h_0-3s^2+3\eud\big)~.
\label{deltah2}
\end{eqnarray}
Note, however, that in the $N_f=4$ theory there are two modular forms of weight 4 that should be considered: the Eisenstein series $E_4$ (that can be multiplied by any modular invariant
term made up with $R$, $N$, $s^2$ and $\eud$) and the following combination
\begin{equation}
 \label{combination2}
T_1 \theta_4^8+2(T_1+T_2)\theta_2^4\theta_4^4+T_2\theta_2^8~,
\end{equation}
whose perturbative limits are, respectively, 1 and $T_1$
Thus, we can easily fix the problem by changing once more the initial condition in the heat-kernel formula and use
\begin{equation}
 \begin{aligned}
  \varphi_0(y;0) \simeq& -\frac{1}{2}\,h_0 
\log\frac{y^2}{\Lambda^2}-\frac{T_1\,\theta_4^4-T_2\,\theta_2^4}{y^2}
-\frac{\Theta_2\,E_4+\Theta_2'\big(T_1 \theta_4^8+2(T_1+T_2)\theta_2^4\theta_4^4+T_2\theta_2^8\big)}{16y^4}~.
 \end{aligned}
\label{init2nf4}
\end{equation}
In this way we obtain the exact expression for $h_2$:
\begin{eqnarray}
h_2 &=&\frac{1}{36}h_0\big(h_0+\eud\big)\big(2h_0+3\eud\big)\,E_2^2
-\frac{4}{3}\big(2h_0+3\eud\big)\big(T_1\,\theta_4^4-T_2\,\theta_2^4\big)\,E_2\nonumber\\
&&+\frac{1}{720}\big(8h_0^3 +68 h_0^2 \eud -48 h_0^2 s^2 +24 h_0 (\eud)^2-48 h_0s^2\eud
-9s^2(\eud)^2+2304 N \big)E_4\nonumber\\
&&-\frac{2}{3}\big(2h_0-3s^2+3\eud\big)\big(T_1 \theta_4^8+2(T_1+T_2)\theta_2^4\theta_4^4+T_2\theta_2^8\big)~.
\label{h2finnf4}
\end{eqnarray}
One can check that (\ref{h1finnf4}) and (\ref{h2finnf4}) precisely match Eq.s~(3.20) and (3.21)
of \cite{Billo:2013fi}. Of course we can continue iteratively this way to determine the
exact expressions of the higher coefficients $h_\ell$'s and hence reconstruct term by term the exact
generalized prepotential. In Appendix~\ref{app:anf4} we provide some details for the calculation
of $h_3$, which shows the consistency and the efficiency of the entire procedure.

\section{S-duality as a Fourier transform}
\label{sft}

The modular anomaly equation is a very powerful tool for several different purposes: in fact
it can be used not only to determine the generalized prepotential $F$, as we have seen in the previous section, but also to investigate its behavior under the S-duality transformation and thus its properties at strong coupling, as we are going to see in this section. In particular, exploiting
the modular anomaly equation or its equivalent heat-kernel version, we will prove the 
following general result: for arbitrary values of $\eu$ and $\ed$, the generalized prepotential $F$ 
and its S-dual $\widetilde F=\cS[F]$ are related through an \emph{exact} Fourier transform, up to 
an important normalization factor, namely
\begin{equation}
\exp\Big(\!\!-\frac{\widetilde F(\tilde a)}{\eud}\Big)\,= \sqrt{\frac{\ii\tau_0}{\eud}}\,
\int_{-\infty}^{+\infty}\!\!dx~\exp\Big(\frac{2\pi\ii\,\tilde a\,x-F(x)}{\eud}\Big)
\label{FT1}
\end{equation}
where $\tilde a = \cS[a]$. 
For simplicity, we will discuss only the $N_f=4$ case, 
but it is obvious that our derivation works in the $\cN=2^*$ theory as well.

Our starting point is the heat-kernel formula (\ref{solnf4}), which for convenience we 
rewrite here as
\begin{equation}
\Psi(a;t) =\big(G * \Psi_0\big)(a;t)
\label{convolution1}
\end{equation}
using the same convolution notation introduced in (\ref{Psi1})-(\ref{convolution}).
We then apply to it an S-duality
transformation with the minimal assumption that%
\footnote{Actually, following the considerations made in \cite{Dimofte:2011jd}
for the $\epsilon$-deformed conformal Chern-Simons theory in three dimensions, one can show
that $\cS$ acts as an exchange of $\eu$ and $\ed$.}
\begin{equation}
\label{Se1e2}
 \cS[\eud]=\eud~,
\end{equation}
and get
\begin{equation}
\exp\Big(\frac{\widetilde \varphi_0(\tilde a;\tilde t)}{\eud}\Big)
=\frac{1}{\sqrt{2\pi\epsilon_1\epsilon_2\tilde t\phantom{\big|}}}
\int_{-\infty}^{+\infty}\!\!d\tilde y ~ 
\exp\Big(\!\!-\frac{(\tilde a-\tilde y)^2}{2\eud \tilde t\phantom{\Big|}}+
\frac{\widetilde \varphi_0(\tilde y;0)}{\eud} \Big)~.
\label{he2}
\end{equation}
Here we have set $\widetilde \varphi_0=\cS[\varphi_0]$, $\tilde y=\cS[y]$ and
\begin{equation} 
\tilde t=\cS[t] =\tau_0^2\Big(t+\frac{1}{2\pi\ii\tau_0}\Big)~.
\label{tildet}
\end{equation} 
Note that this last equation is simply a consequence of the anomalous transformation 
properties of the second Eisenstein series $E_2$ under $\tau_0\to\cS[\tau_0]=-1/\tau_0$. 
Also the expression of $\widetilde\varphi_0(\tilde y;0)$, appearing in the right hand side of
(\ref{he2}), can be easily computed.
In fact, $\varphi_0(y;0)$ is the part of the prepotential with $E_2$ set to zero 
and, up to the logarithmic term, is a sum of terms depending on powers of 
$E_4$, $E_6$ and the Jacobi functions $\theta_j^4$. As one can see from the examples
worked out in the previous section (see, {\it{e.g.}} (\ref{init1nf4}) and (\ref{init2nf4})),
such terms are typically of the form
\begin{equation}
\varphi_0(y;0) \, \propto \, \frac{E_4^\alpha\,E_6^\beta\,(\theta_j^{4})^\gamma\,M^{4\alpha+6\beta+2\gamma+2}}{y^{4\alpha+6\beta+2\gamma}}
\label{term}
\end{equation}
where $\alpha$, $\beta$ and $\gamma$ are non-negative integers and $M$ stands for a generic mass
structure which is needed for dimensional reasons. Taking into account possible exchanges of the 
Jacobi $\theta$-functions among themselves and of the mass invariants $T_1$ and $T_2$
as described in (\ref{modulartheta}) and (\ref{T1T2}), the numerator 
of (\ref{term}) transforms as a modular form of weight $4\alpha+6\beta+2\gamma$, and thus
\begin{equation}
\widetilde\varphi_0(\tilde y;0)\, \propto \,
\tau_0^{4\alpha+6\beta+2\gamma}\,\frac{E_4^\alpha\,E_6^\beta\,(\theta_j^{4})^\gamma\,
M^{4\alpha+6\beta+2\gamma+2}}{\tilde y^{\,4\alpha+6\beta+2\gamma}}~.
\end{equation}
Hence we simply have
\begin{equation}
\widetilde\varphi_0(\tilde y;0) = \varphi_0\Big(\frac{\tilde y}{\tau_0};0\Big)~.
\end{equation}
Plugging this into (\ref{he2}), using (\ref{tildet}) and changing the integration variable
$\tilde y\to\tau_0\,y$, we end up with
\begin{equation}
\label{sHC}
\exp\Big(\frac{\widetilde\varphi_0(\,\tilde a;\tilde t\,)}{\eud}\Big)
=\frac{1}{\sqrt{2\pi\eud\,\tilde t/\tau_0^2\phantom{\big|}}}
~
\int_{-\infty}^{+\infty}\!\!dy~ 
\exp\Bigg(\!-\frac{\big(\frac{\tilde a}{\tau_0}- y\big)^2}{2\eud \,\tilde t/\tau_0^2
\phantom{\Big|}}+
\frac{\varphi_0(y;0))}{\eud} \Bigg)~.
\end{equation}
The right hand side has the same structure as the original equation (\ref{solnf4}); in fact
it is the convolution of the ``initial'' condition $\Psi_0=\exp\big({\frac{\varphi_0(y;0)}{\eud}}\big)$
with a gaussian heat kernel $\widetilde G$ with a rescaled parameter:
\begin{equation}
 \widetilde G(x;\tilde t\,)= \frac{1}{\sqrt{2\pi\eud\,\tilde t/\tau_0^2
\phantom{\big|}}}\,\exp\Bigg(\!-\frac{x^2}{2\eud \,\tilde t/\tau_0^2\phantom{\Big|}}\Bigg)~.
\label{Gtilde}
\end{equation}
Defining 
\begin{equation}
 \Psi_\cS\Big(\frac{\tilde a}{\tau_0};\tilde t\,\Big)
\equiv\exp\Big(\frac{\widetilde\varphi_0(\tilde a;\tilde t\,)}{\eud}\Big)~,
\label{PsiS}
\end{equation}
we can rewrite (\ref{sHC}) more compactly as
\begin{equation}
 \Psi_\cS\Big(\frac{\tilde a}{\tau_0};\tilde t\,\Big) = 
\big(\widetilde G * \Psi_0\big)\Big(\frac{\tilde 
a}{\tau_0};\tilde t\,\Big)~.
\label{convolution2}
\end{equation}
The similarity between (\ref{convolution1}) and (\ref{convolution2}) suggests to take the
Fourier transform $\mathfrak{F}$
\footnote{Our conventions for the Fourier transform $\mathfrak{F}$ and its inverse 
$\mathfrak{F}^{-1}$ are 
\begin{equation*}
\mathfrak{F}[f](k) =  \int_{-\infty}^{+\infty}\!\!dx~ \ee^{-2\pi\ii \,x \,k}\,f(x)
~,~~~ 
\mathfrak{F}^{-1}[g](x) =\int_{-\infty}^{+\infty}\!\!dk~ \ee^{+2\pi\ii\,x\,k}\,g(k)~.
\end{equation*}} 
of both equations. Recalling that the Fourier transform of a convolution of two functions $f$ and $g$
is 
\begin{equation}
\mathfrak{F}[f*g] = \mathfrak{F}[f]~\mathfrak{F}[g]~,
\end{equation}
and that the Fourier transform of a gaussian is
\begin{equation}
\mathfrak{F}\left[\exp(-\alpha x^2)\right](k)
= \sqrt{\frac{\pi}{\alpha}}\, \exp\left(-\frac{\pi^2}{\alpha} k^2\right)~,
\end{equation} 
{from} (\ref{convolution1}) and (\ref{convolution2}) we find
\begin{equation}
\label{FF0}
\begin{aligned} 
&\mathfrak{F}[\Psi](k) =\exp\left(-2\pi^2\eud t\, k^2\right)~\mathfrak{F}[\Psi_0](k)~,
\\
&\mathfrak{F}\left[\Psi_\cS\right] (k) =\exp\Big(\!-\frac{2\pi^2\eud\,\tilde t}{\tau_0^2}\, k^2\Big)~\mathfrak{F}[\Psi_0](k)
\\
&\qquad\qquad~\,=\exp\Big(\!-2\pi^2\eud\big(t+(2\pi\ii\,\tau_0)^{-1}\big)\, k^2\Big)
~\mathfrak{F}[\Psi_0](k)
\end{aligned}
\end{equation}
where in the last step we have used (\ref{tildet}).
By taking the ratio of these two equations, we can eliminate the boundary factor 
$\mathfrak{F}[\Psi_0]$ and also all explicit dependence on $t$, obtaining
\begin{equation}
\label{FF1}
\mathfrak{F}\left[\Psi_\cS\right](k)= 
\exp\left(\frac{\ii \pi\eud}{\tau_0}\, k^2\right)~
\mathfrak{F}[\Psi](k)~.
\end{equation}
Finally, to get $\Psi_{\cS}$ we apply to (\ref{FF1}) the inverse Fourier transform which yields
\begin{equation}
\label{phiss}
\begin{aligned}
\Psi_{\cS}(\tilde x,\tilde t\,) &= 
\int_{-\infty}^{+\infty}\!\!dk~
\ee^{+2\ii\pi k\,\tilde{x}}~
\mathfrak{F}\left[\Psi_\cS\right](k)
\\
&=\int_{-\infty}^{+\infty}\!\!dk~
\ee^{+2\ii\pi k\,\tilde{x}}~\,\ee^{\frac{\ii\pi\eud}{\tau_0}\,k^2}
\int_{-\infty}^{+\infty}\!\!dx~
\ee^{-2\ii\pi k\,x} ~\ee^{\frac{\varphi_0(x;t)}{\eud}} 
\\
&=\sqrt{\frac{\ii \tau_0}{\eud} }~
\ee^{-\frac{\ii \pi \tau_0}{\eud}\,{\tilde x}^2}
\int_{-\infty}^{+\infty}\!\!dx~
\ee^{ \frac{2\pi\ii \tau_0\,\tilde x\,x-\pi\ii\tau_0 x^2+\varphi_0(x,t)}{\eud}}
\end{aligned}
\end{equation}
where the square-root normalization factor in the last line originates from the integral over $k$.

According to (\ref{PsiS}) we must evaluate $\Psi_{\cS}$ at $\tilde x=\tilde a/\tau_0$.
If we do so, we get
\begin{equation}
\exp\Big(\frac{\widetilde\varphi_0(\tilde a;\tilde t\,)}{\eud}\Big) =
\sqrt{\frac{\ii \tau_0}{\eud}}~
\ee^{-\frac{\ii \pi}{\tau_0\,\eud}\,{\tilde a}^2}
\int_{-\infty}^{+\infty}\!\!dx~
\ee^{ \frac{2\pi\ii \,\tilde a\,x-\pi\ii\tau_0 x^2+\varphi_0(x,t)}{\eud}}~,
\end{equation}
which can be put in a more transparent form if we observe that the two terms quadratic in
$x$ and $\tilde a$ appearing in the right hand side are, respectively, the classical parts of 
the prepotential $F$ and of its S-dual $\widetilde F$; indeed
\begin{equation}
\begin{aligned}
&F(x) = \pi\ii\tau_0\, x^2-\varphi_0(x,t)~,\\
&\widetilde F(\tilde a) = \frac{\pi\ii}{\tau_0}\,\tilde a^2- \widetilde\varphi_0(\tilde a;\tilde t\,)~.
\end{aligned}
\label{FSF}
\end{equation}
Taking this into account, we then obtain the announced result (\ref{FT1}), which, for future convenience, we rewrite here after setting $\eud=g_s^2$:
\begin{equation}
\exp\Big(\!\!-\frac{\widetilde F(\tilde a)}{g_s^2}\Big)\,= \sqrt{\frac{\ii\tau_0}{g_s^2}}\,
\int_{-\infty}^{+\infty}\!\!dx~\exp\Big(\frac{2\pi\ii\,\tilde a\,x-F(x)}{g_s^2}\Big)~.
\label{FTfin}
\end{equation}

A few comments are in order. Recently, in \cite{Galakhov:2012gw} it has been argued 
that for generic values of the deformation parameters $\eu$ and $\ed$, the S-duality acts as 
a modified Fourier transform on the prepotential of the $N_f=4$ SYM theory 
for some particular values of the flavor masses, while more recently in \cite{Nemkov:2013qma} 
it has been conjectured, building also on the explicit results of \cite{Billo:2013fi}, that the
S-duality must act precisely as a Fourier transformation. 
Our present analysis provides a general derivation and a proof of this result based on the
modular anomaly equation and the minimal (and natural) assumption (\ref{Se1e2}). 
On the other hand, the fact that the generalized prepotential and its S-dual are related
by an exact Fourier transform is perfectly consistent with the interpretation of
$\exp\left(-\frac{F(a)}{\eud}\right)$ as a wave function, of 
$a$ and $\tilde a=\cS[a]$ as a pair of canonically conjugate variables and, correspondingly,
of the S-duality as a canonical transformation, in complete analogy with what has been observed
in topological string models and local CY compactifications of Type II string theories 
(see for example \cite{Aganagic:2006wq} and references therein). 

Finally, we observe that
by inserting in (\ref{FTfin}) the expansion of $F$ for large values of $x$ 
\begin{equation}
F(x)=\pi\ii\tau_0\,x^2 +\frac{1}{2}\,h_0 \log\frac{x^2}{\Lambda^2}-
\sum_{\ell=1}^\infty \frac{h_\ell}{2^{\ell+1}\,\ell}\,\frac{1}{x^{2\ell}} 
\label{Fx}
\end{equation}
and by carrying out the integration over $x$ using the parabolic cylinder functions as described
in Section~\ref{secn:burger}, we can derive the S-duality transformation properties of the
coefficients $h_\ell$ and check that they indeed are modular forms of weight $2\ell$ with
with anomalous terms due to the presence of the second Eisenstein series $E_2$, in perfect agreement
with the explicit expressions derived from multi-instanton calculations. 

\section{S-duality in the saddle-point approximation}
\label{spa}
In this section we consider the Fourier transform relation (\ref{FTfin}) between the 
generalized prepotential $F$ and its S-dual $\widetilde F$ and evaluate it in the saddle-point approximation for small values of $g_s^2$.
To do so, it is convenient to reorganize the generalized prepotential $F$ in powers of $g_s^2$
and write
\begin{equation}
F=\sum_{g=0}^\infty g_s^{2g}\,F_g~~~~\mbox{with}~~~~
F_g=F_{\mathrm{cl}}\,\delta_{g,0}+\sum_{s=0}^\infty s^{2n}\,F^{(n,g)}~.
\label{Fg}
\end{equation}
As is well-known, the coefficients $F_g$ are related to the refined topological string amplitudes 
at genus $g$. Note that the genus-zero term $F_0$ corresponds to the NS prepotential and that, 
in the limit $s\to 0$, it reduces to $\cF=F_{\mathrm{cl}}+F^{(0,0)}$,
which is the prepotential of the SW theory.

For small $g_s^2$, the integral in (\ref{FTfin}) is dominated by the ``classical'' 
value $x=a_0$ that extremizes the exponent $\big(2\pi \ii\tilde a x-F_0(x)\big)/g_s^2$, 
that is
\begin{equation}
 2\pi\ii \tilde a = \partial F_0(a_0)~.
\label{a0}
\end{equation}
Computing the fluctuations up to second order in $g_s^2$ and using the standard saddle-point method, we 
easily obtain
\begin{equation}
 \widetilde F(\tilde a) =  F(a_0) +g_s^2 \,W_1(a_0) +g_s^4\, W_2(a_0)
-\partial F_0(a_0)\, a_0
\label{SF2}
\end{equation}
where 
\begin{equation}
\begin{aligned}
W_1&= \frac{1}{2}\log\frac{\partial^2\!F_0}{2\pi\ii\tau_0}~,\\
W_2&=\frac{1}{2}\frac{\partial^2\!F_1}{\partial^2\!F_0}
+\frac{1}{8}\frac{\partial^4\!F_0}{{\big(\partial^2\!F_0}\big)^2}
-\frac{1}{2}\frac{\big(\partial F_1\big)^2}{\partial^2\!F_0}
-\frac{1}{2}\frac{\partial F_1 \partial^3\!F_0}{\big(\partial^2\!F_0\big)^2}
-\frac{5}{24}\frac{\big(\partial^3\!F_0\big)^2}{\big(\partial^2\!F_0\big)^3}~,
\end{aligned}
\label{W12}
\end{equation}
and where $a_0$ has to be thought as a function of $\tilde a$ through the inverse of (\ref{a0}).
These expressions agree  with those that can be found for instance in \cite{Aganagic:2006wq}
(modulo the different sign conventions and the different number of variables we are using).
The perturbative corrections $W_g$ have a diagrammatic interpretation in which the propagator is $1/\partial^2 F_0$ and the vertices are given by multiple derivatives of the $F_g$, all evaluated at $a_0$; in the expression of $W_g$ only vertices constructed out of derivatives of $F_k$ with $k<g$ appear; this is again in full similarity to what is described in \cite{Aganagic:2006wq}.

In the NS limit $g_s^2\to 0$, (\ref{SF2}) reduces to
\begin{equation}
 \widetilde F_0(\tilde a) =  F_0(a_0) -\partial F_0(a_0)\, a_0 =
F_0(a_0) -2\pi\ii\tilde a\, a_0
\label{SF0}
\end{equation}
which is the standard and well-known Legendre transform relation between the
NS prepotential and its S-dual. {From} this it is straightforward 
to obtain the S-duality transformations of the derivatives of $F_0$. For
example one has
\begin{align}
&\cS[\partial F_0]=\frac{\partial \cS[F_0]}{\partial \tilde a}=\frac{\partial a_0}{\partial \tilde a}\,\frac{\partial\cS[F_0]}{\partial a_0} 
= -2\pi\ii a_0~,\label{SF0'}\\
&\cS[\partial^2\!F_0] =-\frac{(2\pi\ii)^2}{\partial^2\!F_0}~,
\label{SF0''}
\end{align}
and so on and so forth. Then, exploiting (\ref{SF0'}) and (\ref{a0}), it immediately follows that
\begin{equation}
\label{Ssquareda}
\cS^2[a_0] = \cS[\tilde a]= -a_0~,
\end{equation}
and, using (\ref{SF0}), also that
\begin{equation}
\label{SsquaredF}
\cS^2[F_0] = \cS[\widetilde F_0]= F_0~.
\end{equation}

We now analyze what happens when the corrections in $g_s^2$ are taken into account. We will
show that, with suitable redefinitions, it is possible to write the relation between the prepotential
and its S-dual in the form of a Legendre transform and to generalize the relations
(\ref{Ssquareda}) and (\ref{SsquaredF}) also for finite values of $g_s^2$.
The procedure we follow is similar to the one that is used
to derive the effective action from the generating functional of the connected Green 
functions in quantum field theory, but with some important differences which will point out
later.
Let us consider the first-order corrections in $g_s^2$, namely $F_1$ inside $F$ and
$W_1$ in the S-duality formula. If we want that the right hand side of (\ref{SF2}) becomes
a Legendre transform, we have to redefine the classical saddle-point $a_0$ and the prepotential
$F$ according to
\begin{align}
 a&=a_0+\sum_{g=1}^\infty g_s^{2g}\,\delta a_g~,\label{a}\\
\widehat F&=F + \sum_{g=1}^\infty g_s^{2g} \,\Delta_g  
= F_0 +\sum_{g=1}^\infty g_s^{2g}\big(F_g+\Delta_g\big)~. \label{Fhat}
\end{align}
Inserting these expressions into  (\ref{SF2}), 
and keeping consistently only the terms up to order $g_s^2$, we can rewrite the S-duality relation as
\begin{equation}
 \begin{aligned}
  \cS[\widehat F](\tilde a) =& \,\widehat F(a) -\partial \widehat F(a)\,a  
+g_s^2\Big[W_1(a)-\Delta_1(a)+\cS[\Delta_1](a)\Big]\\
&+g_s^2\Big[\partial^2\!F_0(a)\,\delta a_1+\partial F_1(a)+\partial \Delta_1(a)\Big]a
+\cO(g_s^4)~.
 \end{aligned}
\label{SFhat10}
\end{equation}
This reduces to a Legendre transform if the square brackets vanish, {\it i.e.} if
\begin{align}
 &\Delta_1-\cS[\Delta_1] = W_1~,\label{Delta1}\\
&\delta a_1 =-\frac{1}{\partial^2\!F_0}\big(\partial F_1+\partial \Delta_1\big)~.\label{deltaa1}
\end{align}
It is easy to verify that a solution of these equations is given by
\begin{align}
&\Delta_1= \frac{1}{2}\,W_1= \frac{1}{4}\log\frac{\partial^2\!F_0}{2\pi\ii\tau_0}~,
\label{Delta1sol}\\
&\delta a_1= -\frac{\partial F_1}{\partial^2\!F_0}
-\frac{1}{4}\,\frac{\partial^3\!F_0}{(\partial^2\!F_0)^2}~.
\label{deltaa1sol}
\end{align}
Indeed, from (\ref{SF0''}) and $\cS[\tau_0]=-1/\tau_0$, one has
\begin{equation}
 \cS\Big[\frac{\partial^2\!F_0}{2\pi\ii\tau_0}\Big]=\frac{2\pi\ii\tau_0}{\partial^2\!F_0}~,
\end{equation}
which leads to $\cS[W_1]=-W_1$ and hence to (\ref{Delta1sol}) and, in turn, to (\ref{deltaa1sol}).

This procedure can be extended to higher orders in $g_s^2$ without any difficulty;
in Appendix~\ref{app:c} we provide some details for the computation of the 
corrections at order $g_s^4$, as well as an alternative (and computationally more efficient)
method to determine the higher order terms in the saddle-point expansion. For example, the next-to-leading order correction of the
prepotential turns out to be given by (see (\ref{Delta2solC}))
\begin{equation}
\Delta_2= \frac{1}{4}\frac{\partial^2\!F_1}{\partial^2\!F_0}+
\frac{1}{16}\frac{\partial^4\!F_0}{(\partial^2\!F_0)^2}
-\frac{11}{192}\frac{(\partial^3\!F_0)^2}{(\partial^2\!F_0)^3}~.
\label{Delta2sol}
\end{equation}

This analysis shows that the S-duality relation (\ref{SF2}) can be written as a 
Legendre transform of a
redefined prepotential $\widehat F$, namely
\begin{equation}
 \cS[\widehat F](\tilde a) = \widehat F(a) -2\pi\ii\tilde a\,a
\label{SFhat}
\end{equation}
with
\begin{equation}
\begin{aligned}
\widehat F &=F+\frac{g_s^2}{4}\,\log\frac{\partial^2\!F_0}{2\pi\ii\tau_0}+
g_s^4\Big[\frac{1}{4}\frac{\partial^2\!F_1}{\partial^2\!F_0}+
\frac{1}{16}\frac{\partial^4\!F_0}{(\partial^2\!F_0)^2}
-\frac{11}{192}\frac{(\partial^3\!F_0)^2}{(\partial^2\!F_0)^3}\Big]+\cO(g_s^6)\phantom{\Bigg|}\\
&=F+\frac{g_s^2}{4}\,\log\frac{\partial^2\!F}{2\pi\ii\tau_0}+
\frac{g_s^4}{16}\Big[\frac{\partial^4\!F}{(\partial^2\!F)^2}
-\frac{11}{12}\frac{(\partial^3\!F)^2}{(\partial^2\!F)^3}\Big]+\cO(g_s^6)~,\phantom{\Bigg|}
\end{aligned}
\label{Fhat2}
\end{equation}
and
\begin{equation}
 2\pi\ii\tilde a = \partial\widehat F= \partial F+\frac{g_s^2}{4}\,\frac{\partial^3\!F}{\partial^2\!F}
+\frac{g_s^4}{16}\Big[\frac{\partial^5\!F}{(\partial^2\!F)^2}-
\frac{23}{6}\frac{\partial^3\!F\,\partial^4\!F}{(\partial^2\!F)^3}
+\frac{11}{4}\frac{(\partial^3\!F)^3}{(\partial^2\!F)^4}\Big]+\cO(g_s^6)~.
\label{tildea2}
\end{equation}
Using the formulas derived in Appendix \ref{app:c} (see in particular
(\ref{SF''}-\ref{SF034})), it is easy to check that
\begin{equation}
\label{Ssquared}
\cS^2[a] = \cS[\tilde a]= -a~~~~\mbox{and}~~~~ \cS^2[\widehat F] = \widehat F~,
\end{equation}
which generalize the undeformed S-duality relations (\ref{Ssquareda}) and 
(\ref{SsquaredF}) to the case of a non-vanishing deformation $g_s^2=\eud$. 
Finally, we observe that the expressions for $\widehat F$ and $\tilde a$ we have obtained here completely agree with the results of \cite{Billo:2013fi} that were derived by explicitly 
enforcing, order by order in $\eud$, the requirement that $\cS^2[a]=-a$ 
(see in particular Eq.~(5.3) of \cite{Billo:2013fi} which exactly matches (\ref{Fhat2})).

We end with a few comments. The saddle-point evaluation of the Fourier transform of the deformed prepotential leading to (\ref{SF2}) is a standard procedure, corresponding to the usual perturbative
treatment of quantum field theories. The correspondence is as follows: the topological string coupling $g_s^2 = \eud$ represents $\hbar$; a single degree of freedom $x$ plays the r\^ole of the field $\phi$, while $-2\pi\ii\tilde a$ corresponds to the current $j$; the prepotential $F(x)$ represents 
the tree-level action and its S-dual $\widetilde F(\tilde a)$ is like 
the generating functional $W[j]$ of the connected 
diagrams. In the diagrammatic interpretation of the perturbative corrections $W_g$, like those given 
in (\ref{W12}), $F_0$ is seen as the ``free'' 
action and the higher genus terms $F_g$ as interactions. Note however that these interactions are not weighted by independent couplings, as is usually the case in field theory, but by powers of $g_s^2$, {\it i.e.} of $\hbar$ itself; in other words, one is quantizing a theory whose action already contains 
$\hbar$ corrections. This makes the choice of $F_0$ as the ``free action'' somewhat ambiguous, and 
in fact in Appendix~\ref{app:c} we show that it is possible to organize the expansion around a shifted saddle-point by using as free action a different function, which differs from $F_0$ by 
$\hbar$ corrections, obtaining exactly the same results. 

In the standard field theory procedure one defines the quantum effective action 
$\Gamma[\Phi]$, which is related to $W[j]$ by a Legendre transform, so that $\Phi= \partial W/\partial j=\vev{\phi}_j$. In our situation this would amount to express the
S-dual prepotential as the (inverse) Legendre transform of an ``effective prepotential'' 
$F_{\mathrm{eff}}(a)$, namely as
\begin{equation}
\label{effprep}
\cS[F](\tilde a) = F_{\mathrm{eff}}(a) - 2\pi\ii \tilde a\,a~, 
\end{equation} 
where $a$ can be obtained by inverting%
\footnote{This is tantamount to writing
\begin{equation*}
\label{atrad}
a = -\frac{1}{2\pi\ii} \frac{\partial \widetilde F}{\partial \tilde a} 
= \frac{\int\!dx\, x\, \ee^{\frac{2\pi\ii\tilde a x - F(x)}{g_s^2}}}{ 
\int\! dx\,\ee^{\frac{2\pi\ii\tilde a x - F(x)}{g_s^2}}} = \vev{x}_{\tilde a}
\end{equation*}  
where we used (\ref{FTfin}).} 
the relation  $2\pi\ii\tilde a = \partial F_{\mathrm{eff}}(a)/\partial a$. 

This of course can be done, but it is not what we have done! Indeed, we have modified 
the procedure by introducing a shifted prepotential $\widehat F$ such that its S-dual is the Legendre transform of itself and not of a different function, as one can realize by comparing (\ref{SFhat}) with
(\ref{effprep}).
This result has been obtained by defining the modified prepotential $\widehat F$ to differ 
from $F$ by \emph{half} the perturbative quantum corrections appearing in the 
computation of $\widetilde F(\tilde a)$ (see in particular (\ref{X1sol})\,-\,(\ref{X3sol})).
In other words we have divided the quantum corrections democratically between the prepotential
$\widehat F$ and its S-dual $\cS[\widehat F]$, giving these quantities a symmetric r\^ole with 
respect to S-duality.
In the analogous field theory computation, redefining the action by half the quantum corrections 
would not make much sense, given that one starts from a purely classical action and the goal is precisely to understand how the quantum corrections modify it into an effective action. 
In our case, however, the prepotential $F$ is expressed as a 
series in $\hbar$, {\it{i.e.}} it already contains ``quantum'' terms before the saddle-point 
evaluation of the Fourier transform. It is therefore not particularly disturbing to shift it 
by some $\hbar$ contributions, in order to obtain a better behaved quantity and a more
symmetric formulation of the S-duality relation.

\section{Conclusions}
\label{secn:concl}
The fact that the deformed partition function of $\cN=2$ superconformal $\mathrm{SU}(2)$ theories satisfies a modular anomaly equation, equivalent to the heat equation, was already known in the literature but not all its implications were exploited.

In this paper we have shown that writing the deformed prepotential of $\cN=2^*$ and $N_f=4$ 
theories with gauge group $\mathrm{SU}(2)$ as a solution of the corresponding heat equation is
an efficient way to compute the exact dependence on the bare coupling $\tau_0$ of the coefficients of its large-$a$ expansion. 
With this formalism, quite high orders in the expansion can be achieved with a limited effort, 
using as input data the explicit knowledge of the perturbative (and the very first few instanton) 
terms of the prepotential.

{From} a more conceptual point of view, starting from the solution of the modular anomaly equation 
via the heat kernel, we have proved that $\cS$-duality is realized on the deformed partition function as a Fourier transform also for generic $\epsilon$-deformations and masses. 
This fact is perfectly consistent with the interpretation of
$\exp\left(-\frac{F(a)}{\eud}\right)$ as a wave function and of 
$a$ and $\tilde a=\cS[a]$ as a pair of canonically conjugate variables on which  
S-duality act as a canonical transformation. 
Thus there is a complete analogy with what has been observed
in topological string models and local CY compactifications of Type II string theories 
\cite{Witten:1993ed}\nocite{Aganagic:2006wq}\,-\,\cite{Gunaydin:2006bz}.
The Fourier transform can be evaluated in the saddle-point approximation, yielding the perturbative expansion (\ref{SF2}) of the dual prepotential. As discussed at the end of Section~\ref{spa}, this procedure is very similar to the usual perturbative expansion of field theories, and in particular
it implies that the S-dual variable $\tilde a$ coincides with the derivative of the prepotential $F$ only at the leading order in $g_s^2$. 

In the last part of our work, we have shown that it is possible to introduce a modified prepotential 
$\widehat F$, differing from $F$ by a series of $g_s^2$ corrections determined order by order, 
which exhibits a ``classical'' behavior under S-duality: its S-dual is simply given by its Legendre transform, see (\ref{SFhat}), so that we also have that $\tilde a = \partial \hat F(a)/\partial a$. 
We believe $\widehat F$ should have a direct ``physical'' meaning and uncovering it represents an interesting open problem left to future investigation.
\vskip 1cm
\noindent {\large {\bf Acknowledgments}}
\vskip 0.2cm
This work was supported in part by the MIUR-PRIN contract 2009-KHZKRX and by the Compagnia di San Paolo contract ``Modern Application of String Theory'' (MAST) TO-Call3-2012-0088.

\vskip 1cm

\appendix
\section{Higher order coefficients in the generalized prepotential}
\label{app:a0}
In this appendix we provide some technical details for the calculation of higher
order coefficients $h_\ell$ in the expansion of the generalized prepotential for large values of
the vacuum expectation value $a$. In particular we will compute the coefficient $h_4$ in the
$\cN=2^*$ theory and the coefficient $h_3$ in the $N_f=4$ theory. 
  
\subsection{The coefficient $h_4$ of the $\cN=2^*$ prepotential}
\label{app:a}
The coefficient $h_4^{(0)}$ of the perturbative part of the generalized prepotential
(\ref{phi0p}) for the $\cN=2^*$ SU(2) theory is given by
\begin{eqnarray}
  h_4^{(0)}&=&\frac{1}{1440}h_0\big(h_0+\eud\big)
\big(2h_0^3+13h_0^2\eud-5h_0^2s^2-29h_0\eud s^2+29 h_0 (\eud)^2 +6h_0 s^4\nonumber\\
&&~~~~~\qquad+21 (\eud)^3-42(\eud)^2s^2+21\eud s^4-3s^6\big)~.
\label{h40}
\end{eqnarray}
On the other hand, taking as initial condition for the heat kernel equation the expression
given in (\ref{phi0000}), we obtain a contribution to the prepotential proportional to
$1/a^8$ with a coefficient 
\begin{eqnarray}
 h_4&\!\simeq\!&\frac{1}{20736}\,h_0\big(h_0+\eud\big)
\Big[\big(14h_0^3+79 h_0^2 \eud+ 155h_0(\eud)^2+105(\eud)^3\big)\,E_2^4
\nonumber\\
&&\!\!\!\!+\frac{2}{5}\big(2h_0+3\eud-6s^2\big)\big(14h_0^2+74 h_0\eud+105(\eud)^2\big)\,E_2^2E_4
\label{h4}\\
&&\!\!\!\!+\frac{4}{35}\big(2h_0+7\eud\big)\big(11h_0^2+59h_0\eud+60(\eud)^2-108h_0s^2
-270s^2\eud+180s^4\big)E_2E_6\Big]~.\nonumber
\end{eqnarray}
Notice that this expression does not contain the structure $E_4^2$, which instead
should be expected on the basis of the modularity properties since $h_4$ is a (quasi) 
modular form of weight 8. 
Furthermore, in the perturbative limit when $E_2,E_4,E_6\to 1$, (\ref{h4}) does not
reproduce the 1-loop result (\ref{h40}); indeed we have
\begin{equation}
 \begin{aligned}
  \Theta_4& \equiv h_4^{(0)} - h_4\big|_{E_2,E_4,E_6\to 1} \\
&= 
\frac{1}{725760}h_0\big(h_0+\eud\big)\big(38h_0^3+347h_0^2\eud-480h_0^2s^2+1011 h_0\eud s^2 \\
&~~~~~\qquad+1584 h_0 s^4+ 819(\eud)^3-4789(\eud)^2s^2+5544\eud s^4-1512s^6\big)~.
 \end{aligned}
\label{Theta4}
\end{equation}
This mismatch is easily cured by changing the initial condition in the heat kernel equation (\ref{sol}), and by using
 \begin{equation}
 \varphi_0(y,0) 
\simeq -\frac{1}{2}\,h_0 \log\frac{4y^2}{\Lambda^2} + \frac{\Theta_2\,E_4}{16\,y^4}
+\frac{\Theta_3\,E_6}{48\,y^6}+\frac{\Theta_4\,E_4^2}{128\,y^8}
\label{phi00000}
\end{equation}
instead of (\ref{phi0000}).
In this way, in fact, we obtain the same expressions as before
for $h_1$, $h_2$ and $h_3$, given respectively in (\ref{h11}), (\ref{h22})
and (\ref{h3exact}), and also the exact expression for $h_4$, namely
\begin{eqnarray}
 h_4&\!=\!&\frac{1}{20736}\,h_0\big(h_0+\eud\big)
\Big[\big(14h_0^3+79 h_0^2 \eud+ 155h_0(\eud)^2+105(\eud)^3\big)\,E_2^4
\nonumber\\
&&\!\!\!\!+\frac{2}{5}\big(2h_0+3\eud-6s^2\big)\big(14h_0^2+74 h_0\eud+105(\eud)^2\big)\,E_2^2E_4
\label{h4exact}\\
&&\!\!\!\!+\frac{4}{35}\big(2h_0+7\eud\big)\big(11h_0^2+59h_0\eud+60(\eud)^2-108h_0s^2
-270s^2\eud+180s^4\big)E_2E_6\nonumber\\
&&\!\!\!\!+\frac{1}{35}\big(38h_0^3+347h_0^2\eud-480h_0^2s^2+1011 h_0\eud s^2 \nonumber\\
&&~~~~~\qquad+1584 h_0 s^4+ 819(\eud)^3-4789(\eud)^2s^2+5544\eud s^4-1512s^6\big)E_4^2
\Big]~.\nonumber
\end{eqnarray}
By expanding the Eisenstein series, one may retrieve the various instanton contributions and
one can check that for the first few instanton numbers these contributions exactly agree 
with the explicit results obtained in the multi-instanton calculus 
from the localization formulas \cite{Billo:2013fi}.

\subsection{The coefficient $h_3$ of the $N_f=4$ prepotential}
\label{app:anf4}
In the $N_f=4$ SYM theory the coefficient $h_3^{(0)}$ of the 1-loop prepotential (\ref{f1loopnf4})
is \cite{Billo:2013fi}
\begin{equation}
 \begin{aligned}
 h_3^{(0)} =&~ \frac{1}{672}\Big[24 h_0^4+176 h_0^3\eud-64h_0^3s^2
+372 h_0^2(\eud)^2-352 h_0^2s^2\eud\\
&~~~~~+64 h_0^2s^4-2688 h_0^2 T_1
+124 h_0(\eud)^3-240h_0s^2(\eud)^2+64h_0s^4\eud\\
&~~~~~+3072h_0N-10752 h_0T_1\eud
+4032 h_0T_1s^2 +9216 N\eud-3840Ns^2\\
&~~~~~+15s^4(\eud)^2-36 s^2(\eud)^3
+10080T_1\eud(s^2-\eud)-2016T_1s^4\\
&~~~~~+13056 T_1^2-3072 T_1 T_2-3072 T_2^2\Big]~.
\end{aligned}
\label{h30nf4}
\end{equation}
On the other hand, using the initial condition (\ref{init2nf4})
in the heat-kernel formula (\ref{solnf4}) we obtain the following expression for the coefficient
of the $1/a^6$ term
\begin{eqnarray}
h_3 &\simeq&\frac{1}{216}h_0\big(h_0+\eud\big)\big(5h_0^2+17h_0\eud+15(\eud)^2\big)\,E_3^2\nonumber\\
&&~-\frac{1}{3}\big(5h_0^2+17h_0\eud+15(\eud)^2\big)\big(T_1\,\theta_4^4-T_2\,\theta_2^4\big)
\,E_2^2\nonumber\\
&&~+\frac{1}{1440}\big(2h_0+5\eud\big)\Big[8h_0^3 +68 h_0^2 \eud -48 h_0^2 s^2 +24 h_0 (\eud)^2\label{h3nf4}\\
&&~~~~~\qquad\qquad\qquad\qquad~~-48 h_0s^2\eud-9s^2(\eud)^2+2304 N \Big]E_4E_2\nonumber\\
&&~-\frac{1}{3}\big(2h_0+5\eud\big)\big(2h_0-3s^2+3\eud\big)\big(T_1 \theta_4^8+2(T_1+T_2)\theta_2^4\theta_4^4+T_2\theta_2^8\big)E_2\nonumber\\
&&~+8(T_1\theta_4^4-T_2\theta_2^4)^2 E_2\phantom{\Big|}~.
\nonumber
\end{eqnarray}
It is not difficult to realize that in the perturbative limit where $E_2,E_4,\theta_4\to1$ and 
$\theta_2\to0$ the above expression does not reproduce the 1-loop result (\ref{h30nf4});
indeed
\begin{equation}
 h_3^{(0)}-h_3\big|_{E_4,E_2\to1,\theta_4\to1,\theta_2\to0}=
\Theta_{3} + \Theta_{3}' T_1+16 T_1^2
\label{mismatch}
\end{equation}
where
\begin{eqnarray}
\Theta_3 &=& \frac{1}{30240}\Big[44 h_0^4+1144 h_0^3\eud-864 h_0^3s^2
+4112 h_0^2(\eud)^2-8784 h_0^2s^2\eud\nonumber\\
&&~\qquad+2880 h_0^2s^4+41472 h_0 N+960 h_0(\eud)^3-5382 h_0s^2(\eud)^2\eud\label{Theta30}\\
&&~\qquad+2880 h_0s^4-172800N(s^2-\eud)+675 s^2(\eud)^2(s^2-\eud)\nonumber \\&&~\qquad-138240(T_1^2+T_1T_2+T_2^2)\Big]~,\nonumber\\
\Theta_3'&=&-h_0^2-5h_0\eud+4h_0s^2-3s^4+10s^2\eud-5(\eud)^2~.
\label{Thet31}
\end{eqnarray}
The mismatch (\ref{mismatch}) signals the fact that some structures are missing in our
analysis. In fact, we have three modular forms of weight 6 which have not been considered
so far. They are: 
the Eisenstein series $E_6$, which can be multiplied by any
modular invariant combination of the SO(8) mass invariants (\ref{invdef}), 
and the following two structures:
\begin{equation}
\begin{aligned}
E_4(T_1\theta_4^4-T_2\theta_2^4)~\quad\mbox{and}\quad~
T_1^2(\theta_4^4+2\theta_2^4)\theta_4^8 -
T_2^2(\theta_2^4+2\theta_4^4)\theta_2^8 -T_1T_2\,\theta_2^4\theta_4^4(\theta_2^4-\theta_4^4)
\end{aligned}
 \end{equation}
whose modular transformation properties can be found using
(\ref{modular}) and (\ref{modulartheta}), and whose perturbative limits are, respectively,
$T_1$ and $T_1^2$. Exploiting this fact, we can cure the mismatch (\ref{mismatch}) 
by changing the initial condition in the heat-kernel
formula (\ref{solnf4}) and use
\begin{eqnarray}
 \varphi_0(y;0) &\simeq& -\frac{1}{2}\,h_0 
\log\frac{y^2}{\Lambda^2}-\frac{T_1\,\theta_4^4-T_2\,\theta_2^4}{y^2}
-\frac{\Theta_2\,E_4+\Theta_2'\big(T_1 \theta_4^8+2(T_1+T_2)\theta_2^4\theta_4^4+T_2\theta_2^8\big)}{16y^4}\nonumber\\
&&-\frac{\Theta_3\,E_6+\Theta_3'E_4\big(T_1\,\theta_4^4-T_2\,\theta_2^4\big)}{48y^6}
\label{init3nf4}\\
&&-\frac{16\big(T_1^2(\theta_4^4+2\theta_2^4)\theta_4^8 -
T_2^2(\theta_2^4+2\theta_4^4)\theta_2^8 -T_1T_2\,\theta_2^4\theta_4^4(\theta_2^4-\theta_4^4)\big)}{48y^6}~.
\nonumber
\end{eqnarray}
In this way one obtains the exact expression for $h_3$ which is the sum of (\ref{h3nf4}) and
\begin{equation}
 \Theta_3\,E_6+\Theta_3'E_4\big(T_1\,\theta_4^4-T_2\,\theta_2^4\big)+
16\big(T_1^2(\theta_4^4+2\theta_2^4)\theta_4^8 -
T_2^2(\theta_2^4+2\theta_4^4)\theta_2^8 -T_1T_2\,\theta_2^4\theta_4^4(\theta_2^4-\theta_4^4)\big)~.
\end{equation}
One can check that this result exactly reproduces in all details Eq.(B.2) of \cite{Billo:2013fi},
which was obtained from the multi-instanton calculus and the localization method.

\section{The parabolic cylinder functions}
\label{app:b}
The parabolic cylinder functions $D_q(z)$ have the following integral representation
(see for example 9.241 of \cite{Gradshsteyn})
\begin{equation}
 D_q(z)=\frac{1}{\sqrt{\pi}}\,2^{q+\frac{1}{2}}\,\ee^{-\frac{q\pi\ii}{2}}\,\ee^{\frac{z^2}{4}}\,
\int_{-\infty}^{+\infty}\!du~u^q\,\ee^{-2z^2+2\ii u z}
\label{parabolic}
\end{equation}
for $\mathrm{Re}\,q>-1$, and the following asymptotic expansion for large $z$
(see for example 9.246 of \cite{Gradshsteyn})
\begin{equation}
 D_q(z) \simeq \ee^{-\frac{z^2}{4}}\,z^q\,\sum_{\ell=0}^\infty
\frac{(-1)^\ell\,(q)_{2\ell}}{2^\ell\,\ell!}\,\frac{1}{z^{2\ell}}
\end{equation}
where $\big(q\big)_{2\ell}$ is the Pochhammer symbol
\begin{equation}
 (q)_{2\ell}=q(q-1)\cdots(q-2\ell+1)~.
\end{equation}
Making use of these expressions and their analytic extensions, we can obtain
the useful formula
\begin{equation}
\begin{aligned}
 \int_{-\infty}^{+\infty} dy~y^{q-2k}\,&\exp\Big(\!\!-\frac{(x-y)^2}{2\eud t}\Big)
=\sqrt{2\pi\eud t}~x^{q-2k} \,\sum_{\ell=0}^\infty
\frac{(q-2k)_{2\ell}}{2^\ell\,\ell!}\,\frac{(\eud t)^\ell}{x^{2\ell}}~.
\end{aligned}
\end{equation}
The functions  $D_q(z)$ satisfy the following recursion relations:
\begin{equation}
\begin{aligned}
&D_{q+1}(z)-z D_{q}(z)+q   D_{q-1}(z)=0~,\\
&\partial_z D_{q}(z)+\frac{z}{2}D_{q}(z)-q  D_{q-1}(z)=0~,\\
&\partial_z D_{q}(z)-\frac{z}{2}D_{q}(z)+D_{q+1}(z)=0~,
\end{aligned}
\label{recpara}
\end{equation}
that can be easily rewritten also for the asymptotic series $\widetilde{D}_q(z)$ defined as
\begin{equation}
\widetilde{D}_q(z) =\sum_{\ell=0}^\infty
\frac{(-1)^\ell\,(q)_{2\ell}}{2^\ell\,\ell!}\,\frac{1}{z^{2\ell}} = \ee^{\frac{z^2}{4}}\,z^{-q}\,D_{q}(z)~.
\end{equation}
The functions $\widetilde{D}_q$ are useful since they appear explicitly in the expressions of the various terms 
of the prepotential $\varphi_0(a,t)$ obtained in the main text.
For example, as shown in Section~\ref{secn:burger} for the $\cN=2^*$ SU(2) theory, 
the terms of the prepotential which depend only on the Eisenstein series $E_2$ or which are linear in 
$E_4$ and $E_6$ are reconstructed from the initial condition (\ref{phi0000}), from which
one obtains
\begin{equation}
\varphi_0(a,t) \simeq -\frac{1}{2}\,h_0 \log\frac{4a^2}{\Lambda^2}
+\log \widetilde{D}_q(z) 
+ \frac{\Theta_2\,E_4}{16\,a^4}\frac{\widetilde{D}_{q-4}(z)}{\widetilde{D}_q(z)}
+\frac{\Theta_3\,E_6}{48\,a^6}\frac{\widetilde{D}_{q-6}(z)}{\widetilde{D}_q(z)}+\cdots
\label{phiA2}
\end{equation}
where $t={E_2}/{24}$ and $z={\ii a}/{\sqrt{\eud t}}$, while $\Theta_2$ and $\Theta_3$, which are
respectively polynomials of order $2$ and $3$ in $h_0$, are given in (\ref{Theta2}) and (\ref{Theta3}).

The interesting point is that, while the coefficient of $a^{-2\ell}$ in the expansion of 
$\exp\Big(\frac{\varphi_0(a;t)}{\eud}\Big)$  is a polynomial of degree $2\ell$ in $h_0$ (see (\ref{sol1})), 
the coefficients $h_\ell$ in $\varphi_0$ have to be polynomials in $h_0$ of degree $\ell +1$ for dimensional reasons.
This translates into the fact that while the coefficient of $z^{-2\ell}$ in $\widetilde{D}_q(z)$ is a 
polynomial of degree $2\ell$ in $q$, the coefficients of $z^{-2\ell}$ in
expressions like $\frac{\widetilde{D}_{q-4}(z)}{\widetilde{D}_q(z)}$ or 
$\frac{\widetilde{D}_{q-6}(z)}{\widetilde{D}_q(z)}$ (respectively $\log \widetilde{D}_q(z)$) are polynomials 
of degree $\ell$ (respectively $(\ell+1)$) only. 

We now show that the cancellations needed for this to occur are a simple consequence of the recursion relations 
(\ref{recpara}). We first note that these relations imply 
\begin{equation}
z\partial_z \log \widetilde{D}_{q}(z)=\frac{q(q-1)}{z^2}\,\frac{\widetilde{D}_{q-2}(z)}{\widetilde{D}_{q}(z)}~,
\end{equation}
so that to prove our statement it is enough to demonstrate that for any integer $s$
\begin{equation}
\frac{\widetilde{D}_{q-s}(z)}{\widetilde{D}_{q}(z)}=1 
+\sum_{n\geq 1}\frac{R[q;s;n]}{z^{2n}}
\label{RA}
\end{equation}
where $R[q;s;n]$ is a polynomial in $q$ of order $n$. 

As a lemma, we first demonstrate by induction that, for any integer $s$, we have
\begin{equation}
\frac{\widetilde{D}_{q-s}(z)}{\widetilde{D}_{q}(z)}=1 
+\sum_{n\geq 1}\frac{P[q;s;n]}{z^{2n}}\frac{\widetilde{D}_{q-s-n}(z)}{\widetilde{D}_{q}(z)}
\label{prelRA}
\end{equation}
where $P[q;s;n]$ is a polynomial in $q$ of order $n$. One can easily see that this is true for $s=1$ since
the recursion relations (\ref{recpara}) imply
\begin{equation}
\frac{\widetilde{D}_{q-1}(z)}{\widetilde{D}_{q}(z)}=1 +\frac{q-1}{z^2}\,\frac{\widetilde{D}_{q-2}(z)}{\widetilde{D}_{q}(z)}~.
\label{recA1}
\end{equation}
Then, using the relation (\ref{recA1}) with $q\to q-s$, we get
\begin{equation}
\frac{\widetilde{D}_{q-s-1}(z)}{\widetilde{D}_{q}(z)}=1 
+\sum_{n\geq 1}\frac{P[q;s;n]}{z^{2n}}\frac{\widetilde{D}_{q-s-n}(z)}{\widetilde{D}_{q}(z)}
+\frac{q-s-1}{z^2}\frac{\widetilde{D}_{q-s-2}(z)}{\widetilde{D}_{q}(z)}~,
\label{recs}
\end{equation}
which has the form we seek, except that the index range in the sum is not the correct one. 
However, using (\ref{recA1}) with $q \to q-s-(n+1)$, we can easily put the right hand side of (\ref{recs}) 
in the desired form and thus prove by induction that (\ref{prelRA}) holds true for any integer $s$.
Finally, (\ref{RA}) follows from (\ref{prelRA}) by induction on $n$ and by using the recursion 
(\ref{recA1}) for an appropriate value of $q$.

The functions $\widetilde{D}_{q}$ are useful also for the computation of the prepotential for the $N_f=4$ theory,
which indeed goes essentially along the same lines as above, apart from the existence of an order one term 
proportional to $\big(T_1\,\theta_4^4-T_2\,\theta_2^4\big)$ in the initial condition (\ref{init2nf4}). 
When calculating $h_3$, this term is responsible for the contribution $8(T_1\,\theta_4^4-T_2\,\theta_2^4)E_2$ 
in the last line of (\ref{h3nf4}) which does not seem to have a direct precursor in the initial condition (\ref{init2nf4}). Actually, such a term originates from the fact that the heat equation is satisfied by the partition function rather 
than the prepotential. An explicit computation shows that it is the first term in $\varphi_0 (a,t)$
of a series taking the form
\begin{equation}
\varphi_0 (a,t) \simeq
\frac{(T_1\,\theta_4^4-T_2\,\theta_2^4)^2}{2 \eud \,a^4} 
\Big( \frac{\widetilde{D}_{q-4}(z)}{\widetilde{D}_{q}(z)} - \frac{\widetilde{D}_{q-2}^2(z)}{\widetilde{D}_{q}^2(z)}\Big)+\cdots~,
\end{equation}
which has indeed the features discussed above.

\section{Higher orders in the saddle-point approximation}
\label{app:c}
In this appendix we provide some technical details on the calculation of the next-to-leading
order correction in the S-duality transformation of the prepotential. Before doing this, however,
we make explicit some properties of the leading order results described in Section~\ref{spa} which will
be useful later on.

At order $g_s^2=\eud$, we have found that
\begin{equation}
2\pi\ii\tilde a = \partial F+\frac{g_s^2}{4}\,\frac{\partial^3\!F}{\partial^2\!F}+\cO(g_s^4)=
\partial F_0+g_s^2\Big(\partial F_1+\frac{1}{4}\,\frac{\partial^3\!F_0}{\partial^2\!F_0}\Big)+\cO(g_s^4)~,
\end{equation}
and
\begin{equation}
\widetilde F(\tilde a)=F(a)-2\pi\ii\,\tilde a\,a+\frac{g_s^2}{2}\,\log\frac{\partial^2\!F}{2\pi\ii\tau_0}
+\cO(g_s^4)~.
\end{equation}
{From} these relations it is simple to obtain 
\begin{align}
&\cS[\partial F]=\frac{\partial a}{\partial \tilde a}\,\frac{\partial\cS[F]}{\partial a}
= -2\pi\ii\, a+\frac{2\pi\ii\,g_s^2}{4}\,\frac{\partial^3\!F}{(\partial^2\!F)^2}+\cO(g_s^4)~,
\label{SF'}\\
&\cS[\partial^2\!F] =\frac{\partial a}{\partial \tilde a}\,\frac{\partial\cS[\partial F]}{\partial a}=-\frac{(2\pi\ii)^2}{\partial^2\!F}+\frac{(2\pi\ii)^2 g_s^2}{2}\Big[\frac{\partial^4\!F}{(\partial^2\!F)^3}
-\frac{3}{2}\,\frac{(\partial^3\!F)^2}{(\partial^2\!F)^4}\Big]+\cO(g_s^4)
~,\label{SF''}
\end{align}
and so on and so forth. These relations generalize (\ref{SF0'}) and (\ref{SF0''}) to include the
first-order corrections in $g_s^2$.
Expanding the prepotential as in (\ref{Fg}) and reading the coefficients of $g_s^0$ and $g_s^2$, from (\ref{SF''}) we get
\begin{equation}
\begin{aligned}
 \cS[\partial^2\!F_0] &=-\frac{(2\pi\ii)^2}{\partial^2\!F_0}~,\\
\cS[\partial^2\!F_1] &=(2\pi\ii)^2\Big[\frac{\partial^2\!F_1}{(\partial^2\!F_0)^2}
+\frac{1}{2}\,\frac{\partial^4\!F_0}{(\partial^2\!F_0)^3}-
\frac{3}{4}\,\frac{(\partial^3\!F_0)^2}{(\partial^2\!F_0)^4}\Big]~.
\end{aligned}
\label{SF01''}
\end{equation}
Proceeding similarly for the higher derivatives, we find
\begin{equation}
 \begin{aligned}
&\cS[\partial^3\!F_0]
 =(2\pi\ii)^3\frac{\partial^3\!F_0}{(\partial^2\!F_0)^3}~,
\\
&\cS[\partial^4\!F_0]=(2\pi\ii)^4\,\Big[\frac{\partial^4\!F_0}{(\partial^2\!F_0)^4}
 -3\,\frac{(\partial^3\!F_0)^2}{(\partial^2\!F_0)^5}
  \Big]~.
 \end{aligned}
\label{SF034}
\end{equation}
All these formulas will be useful for the next-to-leading order calculation.

In order to find the shifted prepotential $\widehat F$ at order $g_s^4$, we proceed as described
in Section~\ref{spa} and rewrite the saddle-point result (\ref{SF2}) in terms of
$a$ and $\widehat F$ as given in (\ref{a}) and (\ref{Fhat}), keeping all
terms up to order $g_s^4$. The new structures that in this way are generated in the
right hand side of (\ref{SFhat10}) are
\begin{equation}
\begin{aligned}
 &g_s^4\Big[W_2(a)-\partial W_1(a)\,\delta a_1-\partial F_1(a)\,\delta a_1
-\frac{1}{2}\,\partial^2\!F_0(a)\,\delta a_1^2-\Delta_2(a)+\cS[\Delta_2](a)\Big]\\
&~~+g_s^4\Big[\partial^2\!F_0(a)\,\delta a_2+\partial F_2(a)+\partial \Delta_2(a)
-\frac{1}{2}\,\partial^3\!F_0(a)\,\delta a_1^2\Big]a
\end{aligned}
\label{2ndorder}
 \end{equation}
Thus, in order to have a Legendre transform relation between $\widehat F$ and its S-dual
we must require that the above square brackets vanish. {From} the second line we fix the form
of $\delta a_2$, while from the first line we obtain
\begin{equation}
 \begin{aligned}
 \Delta_2-\cS[\Delta_2] &= W_2-\partial W_1\,\delta a_1-\partial F_1\,\delta a_1
-\frac{1}{2}\,\partial^2\!F_0\,\delta a_1^2\\
&=\frac{1}{2}\frac{\partial^2\!F_1}{\partial^2\!F_0}+
\frac{1}{8}\frac{\partial^4\!F_0}{(\partial^2\!F_0)^2}
-\frac{11}{96}\frac{(\partial^3\!F_0)^2}{(\partial^2\!F_0)^3}
 \end{aligned}
\label{Delta2C}
\end{equation}
where in the second step we have inserted the expressions of $W_1$ and $W_2$ given in
(\ref{W12}) and of $\delta a_1$ given in (\ref{deltaa1sol}). Using (\ref{SF01''}) and (\ref{SF034}),
one can easily check that the combination in the right hand side of (\ref{Delta2C})
changes sign under S-duality; thus a solution to this equation is given by
\begin{equation}
 \Delta_2= -\cS[\Delta_2]=\frac{1}{4}\frac{\partial^2\!F_1}{\partial^2\!F_0}+
\frac{1}{16}\frac{\partial^4\!F_0}{(\partial^2\!F_0)^2}
-\frac{11}{192}\frac{(\partial^3\!F_0)^2}{(\partial^2\!F_0)^3}~,
\label{Delta2solC}
\end{equation}
as reported in the main text.

This procedure can be easily extended to compute higher-order corrections. However, there
is a simpler (and computationally more efficient) way to perform these calculations. Since we aim at
writing the S-duality relation as a Legendre transform for a shifted prepotential $\widehat F$ 
and at having
\begin{equation}
 2\pi\ii \tilde a= \partial \widehat F(a)~,
\label{tildeaC}
\end{equation}
we can organize the calculation in such a way that the identification (\ref{tildeaC}) arises as
the saddle-point condition. Thus, we rewrite (\ref{FTfin}) as
\begin{equation}
\begin{aligned}
 \exp\Big(\!\!-\frac{\widetilde F(\tilde a)}{g_s^2}\Big)\,&= \sqrt{\frac{\ii\tau_0}{g_s^2}}\,
\int_{-\infty}^{+\infty} \!\!dx~\exp\Big(\frac{2\pi\ii\,\tilde a\,x-\widehat F(x)}{g_s^2}\Big)
\,\exp\Big(\frac{\widehat F(x)-F(x)}{g_s^2}\Big)\\
&=\sqrt{\frac{\ii\tau_0}{g_s^2}}\,
\int_{-\infty}^{+\infty} \!\!dx~\exp\Big(\,\frac{2\pi\ii\,\tilde a\,x-\widehat F(x)}{g_s^2}\Big)
\,\exp\Big(\sum_{g=1}^\infty g_s^{2g-2}\,X_{g}(x)\Big)
\end{aligned}
\label{FTC}
\end{equation}
where in the second step we have used
\begin{equation}
 \widehat F= F+\sum_{g=1}^\infty g_s^{2g}\,X_{g}~.
\label{FhatC}
\end{equation}
For small values of $g_s^2$, we can evaluate (\ref{FTC}) in the saddle-point approximation 
by setting 
\begin{equation}
\label{xtoy}
x = a + g_s y
\end{equation}
with $a$ given by (\ref{tildeaC}), so that
\begin{equation}
\label{FT2}
\begin{aligned}
\exp\Big(\!\!-\frac{\widetilde F(\tilde a)}{g_s^2}\Big)  = &
\sqrt{\ii\tau_0}\,\exp\Big(\frac{2\pi\ii\tilde a\,a-F(a)}{g_s^2}\Big) 
\,
\int_{-\infty}^{+\infty} \!\!dy\, 
\exp\Big(\!\!-\frac12 \partial^2\!F(a)\,y^2\Big)\\ 
& \times
\exp\Big(\!\!-\sum_{k=3}^\infty g_s^{k-2}\,
\partial^k\!F(a)\, \frac{y^k}{k!} + y \sum_{g=1}^\infty g_s^{2g-1} \,\partial X_g(a)\Big)~.
\end{aligned}
\end{equation}
We can now expand in powers of $g_s$ and carry out the gaussian integrations, obtaining 
\begin{equation}
\label{FT3}
\exp\Big(\!\!-\frac{\widetilde F(\tilde a)}{g_s^2}\Big)  = 
\sqrt{\frac{2\pi\ii\tau_0}{\partial^2\!F(a)}}\,\exp\Big(\frac{2\pi\ii\tilde a\,a-F(a)}{g_s^2}\Big) 
\Big(1 + \sum_{g=1}^\infty g_s^{2g}\,C_g \Big)~, 
\end{equation}
where the coefficients $C_g$ depend on the quantities indicated below:
\begin{equation}
\label{Cnare}
C_g \equiv C_g\big(\partial X_1,\ldots, \partial X_{g-1};
\partial^2\! F,\ldots, \partial^{2g+2}\!F\big)~.
\end{equation}
Taking the logarithm of (\ref{FT3}), we finally obtain 
\begin{equation}
\label{FT4}
\widetilde F(\tilde a)= F(a) -2\pi\ii  \tilde a\, a  +\sum_{g=1}^\infty g_s^{2g}\,\Gamma_g~,
\end{equation}
where the coefficients $\Gamma_g$ depend on the same type of quantities as the coefficients $C_g$.
The explicit computation of the first few terms yields
\begin{align}
\Gamma_1 =& \frac{1}{2}\log\frac{\partial^2\!F}{2\pi\ii\tau_0}~,
\label{Gamma1}\\
\Gamma_2 =& -\frac{1}{2}\frac{(\partial X_1)^2}{\partial^2\!F} +\frac{1}{2}
\frac{\partial X_1\,\partial^3\!F}{(\partial^2\!F)^2}
+\frac{1}{8}
\frac{\partial^4\!F}{(\partial^2\!F)^2}
-\frac{5}{24}\frac{(\partial^3\!F)^2}{(\partial^2\!F)^3}~,
\label{Gamma2} \\
\Gamma_3 =& -\frac{\partial X_1 \partial X_2}{\partial^2\!F} 
+\frac{1}{2}\frac{\partial X_2\,\partial^3\!F}{(\partial^2\!F)^2} +
\frac{1}{6}\frac{(\partial X_1)^3\,\partial^3\!F}{(\partial^2\!F)^3}
+\frac{1}{4} \frac{(\partial X_1)^2\partial^4\!F}{(\partial^2\!F)^3} + 
\frac{1}{8}\frac{\partial X_1\,\partial^5 \!F}{(\partial^2\!F)^3} \notag \\
&+
\frac{1}{48}\frac{\partial^6\!F}{(\partial^2\!F)^3}
-\frac{1}{2}\frac{(\partial X_1)^2\,(\partial^3\!F)^2}{(\partial^2\!F)^4} -\frac{2}{3}
\frac{\partial X_1\, \partial^3\!F\,\partial^4\!F}{(\partial^2\!F)^4} 
-\frac{7}{48} \frac{\partial^3\!F\,\partial^5\!F }{(\partial^2\!F)^4}
-\frac{1}{12}\frac{(\partial^4\!F)^2}{(\partial^2\!F)^4}\notag \\
&+\frac{5}{8} \frac{\partial X_1\,(\partial^3\!F)^3}{(\partial^2\!F)^5} + \frac{25}{48}\frac{(\partial^3\!F)^2\,\partial^4\!F}{(\partial^2\!F)^5}
- \frac{5}{16}\frac{(\partial^3\!F)^4}{(\partial^2\!F)^6}~.
\label{Gamma3}
\end{align}
Obtaining higher $\Gamma_g$'s is quite straightforward.

Using (\ref{FhatC}), we can rewrite (\ref{FT4}) as
\begin{equation}
\label{FT5}
\cS[\widehat F](\tilde a)= \widehat F(a) -2\pi\ii  \tilde a\, a  +\sum_{g=1}^\infty g_s^{2g}\,
\Big[\cS(X_g)-X_g+\Gamma_g\Big]~,
\end{equation}
and hence we have the desired Legendre transform relation if
\begin{equation}
 \cS[X_g]-X_g+\Gamma_g=0~.
\label{conditions}
\end{equation}
These equations can be easily solved iteratively. For example, we have
\begin{equation}
 X_1=-\cS[X_1] = \frac{1}{2}\,\Gamma_1=\frac{1}{4}\log\frac{\partial^2\!F}{2\pi\ii\tau_0}~;
\label{X1sol}
\end{equation}
plugging this expression into (\ref{Gamma2}), we then obtain
\begin{equation}
 X_2=-\cS[X_2] = \frac{1}{2}\,\Gamma_2=\frac{1}{16}\frac{\partial^4\!F}{(\partial^2\!F)^2}
-\frac{11}{192}\frac{(\partial^3\!F)^2}{(\partial^2\!F)^3}~;
\label{X2sol}
\end{equation}
in turn, using these results into (\ref{Gamma3}), we find
\begin{equation}
\label{X3sol}
\begin{aligned}
 X_3=-\cS[X_3] =\frac{1}{2}\,\Gamma_3=& ~\frac{1}{96}\frac{\partial^6\!F}{(\partial^2\!F)^3} - 
\frac{19}{384}\frac{\partial^3\!F\,\partial^5\!F }{(\partial^2\!F)^4}
-\frac{1}{24}\frac{(\partial^4\!F)^2}{(\partial^2\!F)^4}\\
&+ \frac{119}{768}\frac{\partial^4\!F\,(\partial^3\!F)^2}{(\partial^2\!F)^5} + 
\frac{109}{1536}\frac{(\partial^3\!F)^4}{(\partial^2\!F)^6}~,
\end{aligned}
\end{equation}
and we can continue iteratively this way. The prepotential $\widehat F=F+g_s^2\,\Gamma_1+g_s^4\,\Gamma_2+\cdots$
perfectly agrees with the one derived with the more ``conservative'' saddle-point method
as described in the main text and at the beginning of this appendix (see in particular
(\ref{Fhat2})).

\providecommand{\href}[2]{#2}\begingroup\raggedright
\endgroup

\end{document}